\documentclass[a4paper,11pt]{article}

\usepackage{jheppub} 

\usepackage[T1]{fontenc} 
\usepackage{natbib}
\usepackage{setspace}

\def\be{\begin{equation}}
\def\ee{\end{equation}}

\def\bea{\begin{eqnarray}}
\def\eea{\end{eqnarray}}

\def\vec[#1]{\boldsymbol{#1}}
\def\vecs[#1,#2]{\boldsymbol{{#1}_{#2}}}

\def\mes[#1]{d^{3}{#1}}
\newcommand{\vect}[1]{{\boldsymbol{#1}}}
\def\del{\partial}

\def\vk{\vect{k_1}}
\def\vkk{\vect{k_2}}
\def\vkkk{\vect{k_3}}
\def\vkfour{\vect{k_4}}
\newcommand{\half}{\frac{1}{2}}
\newcommand{\quarter}{\frac{1}{4}}

\title{Constraints from Conformal Symmetry on the Three Point Scalar Correlator in Inflation}

\author[a]{Nilay Kundu,}
\author[b]{Ashish Shukla,}
\author[b]{and Sandip P. Trivedi}

\vspace{5mm}

\affiliation[a] {\it Harish-Chandra Research Institute, Chhatnag Road, Jhunsi,
Allahabad, 211019, India}
\affiliation[b]
{\it Department of Theoretical Physics,
 Tata Institute of Fundamental Research,\\  Colaba, Mumbai, 400005, India \\}

\vspace{1cm}

\emailAdd{nilay.tifr@gmail.com}
\emailAdd{ashukla.phy@gmail.com}
\emailAdd{trivedi.sp@gmail.com}

\abstract{Using symmetry considerations, we derive Ward identities which relate the three point function of scalar perturbations produced during inflation to the scalar four point function, in a particular limit. The derivation assumes approximate conformal invariance, and the conditions for the slow roll approximation, but is otherwise model independent. The Ward identities  allow us to deduce that the three point function must be suppressed in general, being  of the same  order of magnitude as in the slow roll model. They also  fix the three point function  in terms of the four point function, upto one constant which we argue is generically suppressed. Our approach is based on analyzing the wave function of the universe, and the Ward identities arise by imposing the requirements of spatial and time reparametrization invariance on it.}

\preprint{\parbox{3cm}{TIFR/TH/14-24 
 \\ HRI/ST/1414
}}

\begin{document} 
\maketitle
\flushbottom

\section{Introduction}
\label{intro}
Inflation is an attractive idea which explains the approximate homogeneity and 
isotropy of our universe. It also  leads to the generation of small 
perturbations required for the observed anisotropy in the Cosmic Microwave 
Background and for the growth of structure. 

Despite considerable attention having been devoted to this idea theoretically, 
relatively little work has been done on understanding the nature of the 
perturbations which are produced during inflation, in a model independent 
manner. More recently, such a model independent analysis has been developed 
using symmetry considerations.
 
During inflation, spacetime  is approximately  described by de Sitter space. 
The essential idea of some of  the symmetry based analysis is to use the $SO(4,1)$ symmetry of de Sitter space, which is 
also the symmetry group of three dimensional Euclidean Conformal Field 
Theories, to constrain correlation functions of the perturbations. Of course, the 
universe is not exactly described by de Sitter space during inflation, but the 
corrections which are quantified in terms of the slow roll parameters are 
small, being of order 1\% or so.  The $SO(4,1)$ symmetry should therefore be useful in constraining the  correlation functions.
 In the discussion below, we 
shall refer to this symmetry group as the de Sitter group or the 
conformal group interchangeably.

In this paper  we carry out such a symmetry based analysis for the scalar three point function, by including the first non-vanishing corrections in the slow roll approximation. 
Among all the three point correlations, the three point scalar  correlator  is 
expected to be of the biggest magnitude, and therefore of most significance for 
observational tests of non-Gaussianity. It is therefore clearly important to 
understand what constraints can be imposed on it from symmetry considerations 
alone.  This is the motivation underlying our work.

One of our main results is a set of Ward identities  relating the three point function to the scalar four point function in a particular limit. 
The coefficient of proportionality between the two is the parameter ${\dot{\bar{\phi}}\over H}$, defined in section \ref{basics}. 

It is well known that the three point function is suppressed in the  canonical model of slow roll 
 inflation (for a  definition of this model see eq.(\ref{actslroll})) so that, in a sense which we make precise below, it can be 
thought of as  vanishing to  leading order in the slow roll approximation. We argue that this feature is more generally valid. 
In addition, the Ward identities  allow us to  estimate the magnitude of the leading non-vanishing contribution to the three point function, in the slow roll approximation. 
We find that generically it is of the same order as the three point function in the canonical slow roll model. To get a rough idea, this means that 
quite generally, as long as conformal symmetry is approximately valid, $f_{NL}\sim O\big(({\dot{\bar{\phi}}\over H})^2\big)$, although the detailed functional form is not  the same as  assumed in the standard $f_{NL}$ parametrization, so this is only an estimate. 

While the small magnitude for the three point function is  disappointing from the point of view of observations, this result 
can be turned around in an interesting way  as follows. If observationally a 
three point function of bigger magnitude is observed then it would rule out not only  the canonical model of slow roll inflation, but in fact all models where the dynamics is 
approximately conformally invariant, and the slow roll approximation holds \footnote{Strictly speaking, in a non-generic case,  approximate conformal invariance and the slow roll approximation do allow the magnitude to be bigger, as we discuss below. But in this case the functional form is completely fixed, so one should be able to test for this possibility as well.}. 

We also show that the Ward identities  determine the three point function, nearly completely, upto one constant, in terms of the four point function. To  leading order, the latter can be 
computed in the de Sitter limit and is thus constrained by the full de Sitter 
symmetry group. In this way, we can make precise the extent to which conformal 
symmetry constrains the scalar three point correlator. 

Unfortunately, as is well known, the four point function itself is not 
significantly constrained in a conformal field theory. 
 In position space there are three invariant cross ratios in  three dimensions, 
and conformal symmetry allows the four point scalar correlator   to  be a 
general function of these three variables. This is a rather weak constraint. It 
follows from our analysis then that conformal invariance  also constrains the 
three point scalar correlator  only weakly. 

Directly checking the Ward identities through observations seems very challenging, 
although it cannot be ruled out, perhaps.  A more interesting  angle might be 
the following. 
In the canonical slow roll model, the four point function in the de Sitter limit arises from a tree diagram with single graviton exchange, see \cite{Seery:2008ax}, \cite{Ghosh:2014kba}. If the three point 
function is observed and found to depart from the functional form it has in the canonical
slow roll model, then it would follow from the Ward identities  that 
the four point function must also have a different form. This would suggest that perhaps 
higher spin fields must have been involved during inflation. We leave a study along these lines for the future.

The approach we follow  in this paper is based on the important work of \cite{Maldacena:2002vr} and \cite{Maldacena:2011nz} and also the subsequent papers, \cite{Mata:2012bx} and \cite{Ghosh:2014kba}.  
As was emphasized in  these works,  symmetry considerations  are conveniently discussed in terms of 
 the  wave function of the universe at late times. In the de Sitter 
limit, the Ward identities of conformal invariance can be obtained from the 
constraints of spatial reparametrization and time reparametrization invariance, 
which the wave function must satisfy. The time reparametrization constraint in 
particular is the same as the Wheeler-DeWitt equation.  These constraints must continue to hold even when we go beyond the de Sitter limit.
In this way, the spatial and time reparametrization invariance can be  used  to obtain the  corrected Ward identities which now include the breaking of conformal invariance.

Some of the Ward identities we obtain have already been discussed in  the literature, see for example \cite{Creminelli:2012ed} for an early discussion. 
Part of our motivation  in presenting them  here is to show that they follow from 
the more  general approach mentioned  in the previous paragraph. In this paper we have only analyzed the three point scalar correlator, and that too to leading  non vanishing order for which  the analysis is relatively  straightforward. But in principle this approach should be extendable for all correlators
order by order in the slow roll expansion. 
 We leave a general analysis of this kind for the future.

It is  worth explicitly mentioning that while the analysis we carry out draws on techniques developed in the study of the AdS/CFT correspondence, we do not  assume that there is a hologram for de Sitter space or for inflation. 
We use the techniques drawn from AdS/CFT only  as a way of efficiently organizing the analysis of symmetry constraints
for  perturbations which are generated during inflation in the gravitational system.

The  analysis  we carry out assumes, as was mentioned above, that the full inflationary dynamics, including the scalar sector,
preserves approximate conformal invariance. Our conclusions therefore do not apply to models like DBI inflation \cite{Silverstein:2003hf, Alishahiha:2004eh} or Ghost inflation \cite{ArkaniHamed:2003uz}, in which the scalar sector breaks  the full conformal symmetry badly.
In addition, it  assumes  that  only one inflaton was present during inflation,   and  that the initial state was  the Bunch-Davies vacuum. We also assume that the slow-roll conditions hold; these  are  more precisely discussed in section \ref{moregenaction}. 
 Besides these assumptions, our conclusions are robust, and as was emphasized above, model independent. For example, they should hold even if higher derivative corrections to Einstein gravity become important \footnote{More correctly, these results should   apply  also to  models where quantum effects are small but classical higher derivative corrections are important.  As would   happen, for example, if the Hubble scale is of order the string scale, $M_{st}$, but much  smaller than the Planck scale, $M_{Pl}$.}.

This paper is organized as follows. In section \ref{basics} we discuss some of the introductory material. The Ward identities are derived in section \ref{wardderiv}.
In section \ref{commentward} we analyze these identities further and derive various consequences. Finally, we conclude  in section \ref{conclusions}. 
Appendices \ref{appcoeffn}, \ref{appSh} and \ref{prescription} contain additional important details. 

Before concluding we should discuss  some of the related literature. 
Early work on using conformal symmetry to constrain inflationary correlators includes \cite{Antoniadis:1996dj, Larsen:2002et, Larsen:2003pf, McFadden:2010vh, Antoniadis:2011ib, McFadden:2011kk, Creminelli:2011mw, Bzowski:2011ab, Kehagias:2012pd, Kehagias:2012td, Schalm:2012pi, Bzowski:2012ih}. 
More recent work, where the conformal symmetries are often thought of as being non-linearly realized, include \cite{Weinberg:2003sw, Weinberg:2008nf, Creminelli:2004yq, Cheung:2007sv, Senatore:2012wy, Creminelli:2011sq, 2012JCAP02017B, Creminelli:2012ed, Creminelli:2012qr, Hinterbichler:2012nm, Assassi:2012zq, Goldberger:2013rsa, Hinterbichler:2013dpa, Creminelli:2013cga, Berezhiani:2013ewa, Sreenath:2014nka, Mirbabayi:2014zpa, Joyce:2014aqa, Sreenath:2014nca}. 
Many interesting Ward identities have already been derived using this approach. 
Additional related work is in \cite{Banks:2011qf, Banks:2013qra}, see also \cite{Pimentel:2013gza}. 
Our discussion in section \ref{commentward} is closely related to \cite{Bzowski:2013sza}, see also \cite{Coriano:2013jba}. 
The basic approach of using time and spatial reparametrizations to derive Ward identities that we follow was first discussed in the AdS context in \cite{deBoer:1999xf}. More recently, there are   related developments in the study of Lifshitz and hyperscaling violating spacetimes, of interest for  possible connections between AdS gravity  and condensed matter physics, see \cite{Chemissany:2014xsa}.


\section{Basic Set Up and Conventions}
\label{basics}
Here we give a few details about the basic approach we will use, for more 
details see \cite{Maldacena:2002vr}, and \cite{Mata:2012bx}, \cite{Ghosh:2014kba}. 

We will consider the metric to be of the ADM form 
\be
\label{metadm}
ds^2=-\,N^2 dt^2 + h_{ij} \, (dx^i + N^i dt) (dx^j+N^j dt),
\ee

and work in the gauge 
\be
\label{gaugec}N=1, N^i=0.
\ee
The equations of motion obtained by varying $N$ and $N^i$ in the action must 
still be imposed. These equations will give rise to the constraints of spatial 
and time reparametrizations that  play an important role in the subsequent 
discussion. 

The background inflationary solution is a Friedmann-Robertson-Walker (FRW) spacetime with scale factor 
$a(t)$. Allowing for perturbations in the metric, we can write
\be
\label{metrpert}
h_{ij} \equiv a^2(t) \, g_{ij} = a^2(t) \, [\delta_{ij} +\gamma_{ij}],
\ee
with 
\be
\label{defzeta}
\gamma_{ij} = 2 \zeta \delta _{ij} + \widehat{\gamma}_{ij}
\ee
where $\widehat{\gamma}_{ij}$ is traceless. 

A scalar field, the inflaton, $\phi$,  is also present in inflation (as 
mentioned in the introduction, we will restrict ourselves to the case with a single 
inflaton). It can be written as 
\be
\label{infla}
\phi = {\bar \phi}(t) + \delta \phi(t,\vec[x])
\ee
where ${\bar \phi}$ and $\delta \phi$ are the background value and the 
perturbation of the inflaton, respectively.

We will consider the wave function of the universe at late times, when the 
perturbations of interest have exited the horizon and stopped evolving in time. 
The wave function is actually a functional of the perturbations $\gamma_{ij}, 
\delta \phi$.
Assuming the wave function is approximately Gaussian and that corrections are 
small, we can expand it in a Taylor series in the perturbations to get 
\begin{align}
\label{wf1}
\begin{split}
\Psi[\delta \phi, \gamma_{ij}] = \text{exp} \bigg[ &\frac{M_{Pl}^{2}}{H^2} \bigg(- \half \int 
\mes[x] \sqrt{g(\vec[x])} \, \mes[y] \sqrt{g(\vec[y])} \, \delta\phi(\vec[x]) \, \delta\phi(\vec[y]) \langle 
O(\vec[x]) O(\vec[y]) \rangle
\\ &- \half \int \mes[x] \sqrt{g(\vec[x])} \, \mes[y] \sqrt{g(\vec[y])} \, \gamma_{ij}(\vec[x]) \gamma_{kl}(\vec[y]) \langle T^{ij}(\vec[x]) T^{kl}(\vec[y]) \rangle
\\ &+ \frac{1}{3!} \int \mes[x] \sqrt{g(\vec[x])} \, \mes[y] \sqrt{g(\vec[y])} \, \mes[z] \sqrt{g(\vec[z])} 
\\ & \hspace{37mm} \delta\phi(\vec[x]) \, \delta\phi(\vec[y]) \, \delta\phi(\vec[z]) \langle O(\vec[x]) O(\vec[y]) O(\vec[z]) \rangle 
\\ &+ \frac{1}{4!}  \int \mes[x] \sqrt{g(\vec[x])}\, \mes[y] \sqrt{g(\vec[y])}\, \mes[z] \sqrt{g(\vec[z])}\, \mes[w] \sqrt{g(\vec[w])}
\\ & \hspace{15mm} \delta\phi(\vec[x]) \, \delta\phi(\vec[y]) \, \delta\phi(\vec[z]) \, 
\delta\phi(\vec[w]) \langle O(\vec[x]) O(\vec[y]) O(\vec[z]) O(\vec[w]) \rangle + \ldots \bigg) \bigg].
\end{split}
\end{align}
The ellipses denote additional terms which will not play an important role in this paper.

The coefficient function for the quadratic term in $\delta \phi$ in eq.(\ref{wf1}) is given by \footnote{We follow the convention where bold face symbols, e.g. $\vec[k]$, stand for 3-vectors, and symbols without bold faces denote the magnitudes, e.g. $k \equiv |\vec[k]|$.}
\be
\label{defoo}
\langle O(\vec[k]) O(\vec[k']) \rangle=(2\pi)^3 \delta^3(\vec[k]+\vec[k']) \, k^3.
\ee
Let us also mention that in our conventions
\be
\label{ftcon}
\langle O(\vec[k]) O(\vec[k']) \rangle = \int d^3x \, d^3y \, e^{-i \vec[k]\cdot \vec[x]} \, e^{-i \vec[k']\cdot \vec[y]} \langle O(\vec[x]) O(\vec[y]) \rangle.
\ee
We also note that the coefficient function for the quadratic term in $\gamma_{ij}$ is given by \footnote{The labels $s, s'$ denote the two polarizations of the graviton.}
\be
\label{deftt}
\langle T^s(\vecs[k,1]) \, T^{s'}(\vecs[k,2]) \rangle = (2\pi)^3 \delta^3(\vecs[k,1]+\vecs[k,2]) \, \delta^{s,s'} \, \frac{k_1^{\,3}}{2} ,
\ee
where $T^s(\vec[k]) = T_{ij}(\vec[k]) \, \epsilon^{s,ij}(-\vec[k])$,
and the polarization tensor, $\epsilon^{s,ij}$, satisfies the normalization $\epsilon^{s,ij} \epsilon^{s'}_{\, ij} = 2 \, \delta^{s,s'}$.

The wave function eq.(\ref{wf1})  is obtained by doing a path integral with Bunch-Davies 
boundary conditions in the far past,
\be
\label{wf2}
\Psi[\delta \phi, \gamma_{ij}]                                                                                                                                                                                                                                                                                                                                                                                                                                                                                                                                       = \int [\mathcal{D} \delta\phi] \, [\mathcal{D}\gamma_{ij}] \, e^{i \,S[\delta \phi, \gamma_{ij}] }\,.
\ee

Our choice, eq.(\ref{gaugec}), does not fix the gauge completely. There is still 
the freedom to do spatial reparametrizations of the form
\be
\label{spacere}
x^i\rightarrow x^i + \epsilon^i(\vec[x]) ,
\ee
and time reparametrization of the form
\be
\label{timere}
t \rightarrow t + \epsilon(\vec[x]) , \, x^i \rightarrow x^i + v^i(t,\vec[x]), 
\ee
where
\be
\label{videff}
v^i=\partial_i\epsilon \int {1\over a^2(t) } \, dt.
\ee
Note that in de Sitter space eq.(\ref{videff}) becomes,
\be
\label{videf}
v^{i} = - \, \frac{1}{2H} \, (\partial_{i}\epsilon) \, e^{-2Ht}.
\ee

The wave function must be  invariant under these coordinate transformations.
In the classical limit, which we mainly consider here, the wave function is 
approximately
\be
\label{approxwf}
\Psi[\delta \phi, \gamma_{ij}]  \sim e^{i\, S[\delta \phi, \gamma_{ij}]},
\ee
and the invariance of the wave function arises from the invariance of the action 
with respect to the spatial and time reparametrizations. 
It is easy to see in  the Hamilton-Jacobi formulation that for Einstein gravity, 
for example, the equation of motion obtained by varying $N, N^i$ in the action, are 
exactly the equations which impose this invariance. More generally, the 
equations of motion can be complicated, but the ones obtained by varying $N, N^i$ 
should, on general grounds,  still impose this invariance. 

The invariance of the wave function under eq.(\ref{spacere}) and 
eq.(\ref{timere}) leads to conditions on the coefficient functions, introduced 
in eq.(\ref{wf1}). 
In de Sitter space these constraints are exactly the same as Ward identities for 
conformal invariance in a conformal field theory, with the coefficient 
functions playing the role of correlation functions in the CFT. This is the 
essential reason why the study of the constraints imposed by conformal 
invariance on the wave function, and therefore expectation values, can be mapped 
to an analysis of constraints imposed on correlation functions in a CFT.

In de Sitter space the scale factor, eq.(\ref{metrpert}). is given by 
\be
\label{scfactor}
a^2(t)=e^{2 H t}
\ee
where $H$, the Hubble parameter, is constant. 
More generally the Hubble parameter, defined by 
\be
\label{defH}
\frac{\dot{a}}{a}\equiv H
\ee
will not be a constant. 

Its variation gives two of the slow roll parameters which quantify the breaking 
of conformal invariance,
\be
\label{defslowroll}
\epsilon = - \, \frac{\dot{H}}{H^2}\,,~ \delta = \frac{\ddot{H}}{2H\dot{H}} \,.
\ee
Another parameter is given by 
\be
\label{para2}
\frac{\dot{\bar{\phi}}}{H} \, .
\ee

We often refer in this paper to the ``canonical model of slow roll inflation''. By this we mean a theory with   the action 
\be
\label{actslroll}
S=\int d^4x \, \sqrt{-g} \, M_{Pl}^2  \bigg[\half R - \half (\nabla \phi)^2   
- V(\phi) \bigg],
\ee
where the potential is varying slowly enough to meet the conditions, eq.(\ref{condslra}) and eq.(\ref{condbsl}). 
Note that in our normalization the scalar  field is dimensionless, and $V$ has dimensions of $[M]^2$.
In this theory the Hubble parameter is given by 
\be
\label{hubble}
H^2 = {1\over 3} \, V.
\ee

In the slow roll approximation in this model, the conditions 
\be
\label{condslra}
\epsilon, \delta \ll 1,
\ee
and also
\be
\label{condbsl}
{\dot{ \bar{\phi}}\over H} \ll 1,
\ee
are met.

The scalar field then approximately satisfies the equation
\be
\label{sceq}
\dot{\bar{\phi}} \simeq - \, {1\over 3 H} \, V' ,
\ee
where a prime denotes derivative with respect to the scalar field. The slow roll parameters, $\epsilon$ and  $\delta$, defined in eq.(\ref{defslowroll}), are  given by 
\be
\label{epsdelslr}
\epsilon={1\over 2} \left({V'\over V}\right)^2 ~~ \text{and} ~~ \delta = \epsilon - {V''\over V},
\ee
 and meeting the slow roll conditions, eq.(\ref{condslra}), eq.(\ref{condbsl}) leads to the requirements,
\be
\label{condpotsl}
\left({V'\over V}\right)^2\ll 1,
\ee
and
\be
\label{condpotalb}
{V''\over V} \ll 1.
\ee

Also,  in this model
\be
\label{relasl}
\frac{\dot{\bar{\phi}}}{H}=\sqrt{2 \epsilon}.
\ee
As a result, from eq.(\ref{condslra}) we see that 
\be
\label{conddd}
{\dot{\bar{\phi}}\over H} \gg \epsilon, \delta.
\ee

\subsection{More General Action and Slow Roll Conditions}
\label{moregenaction}
More generally, our analysis will allow for additional terms so that the full action could schematically take the form
\be
\label{genlag}
S= \int d^4x \, {\sqrt{-g}} \, M_{Pl}^2 \, \bigg[\half R - {1\over 2} (\partial \phi)^2 - V+ {c_1 \over \Lambda^2}R^2 + {c_2 \over \Lambda^4} R^3 + \cdots \bigg],
\ee
where the  additional terms, like the last two, also have additional derivatives.  The  $R^2, R^3$  terms above actually denote various terms with four derivative and six derivatives respectively.  The coefficients $c_1, c_2$ are dimensionless,
and in general could be functions of $\phi$, while  $\Lambda$ denotes a higher energy cut-off scale, which could in string theory  be the string scale, $M_{st}$, for example.  
The $R^2, R^3$  terms could be significant, for example, if the Hubble scale is of order the string scale in string theory. The ellipses stand for additional terms  with higher derivatives on the metric, and also terms with additional derivatives on the inflaton. These would be suppressed by appropriate powers of  $\Lambda$.

As was mentioned above, we are interested here in theories where the additional terms in eq.(\ref{genlag})  give rise to an approximately conformally invariant dynamics  for the perturbations. This can be ensured by taking both the Hubble parameter and the scalar to vary slowly, so that eq.(\ref{condslra}) and eq.(\ref{condbsl}) are met.
The background solution is then approximately de Sitter space with a constant scalar, which clearly preserves conformal invariance. And the perturbations about this background will then inherit this conformal symmetry.
In the discussion which follows, it will be convenient for parameter counting to take  
\be
\label{slcomp}
\epsilon \sim \delta.
\ee 
Corrections about the conformally invariant limit will then be suppressed by $\epsilon$ and ${\dot{\bar{\phi}} \over H}$. 
With these features in mind we will take, in general,  the conditions eq.(\ref{condslra}) and eq.(\ref{condbsl}) to hold 
for approximate conformal invariance to arise\footnote{ Strictly speaking, we have established that the conditions eq.(\ref{condslra}), eq.(\ref{condbsl}) are sufficient, but perhaps not necessary. However, if they are violated the emergence of approximate conformal invariance for the dynamics of small perturbations would be something of an accident, which we view as being quite unlikely.}.

Once these conditions are met, it also follows from the field equations in the general case that eq.(\ref{conddd}) is  valid. As was mentioned above, we are assuming that there is  an approximate de Sitter solution  when ${\dot{\bar{\phi}}\over H} $ is small. The corrections to de Sitter space in such a solution  arise because of extra contributions to the stress energy
 due to  the non-vanishing value of $\dot{\bar{\phi}}$. However, any such contribution must be of order $(\dot{\bar{\phi}})^{2}$ or higher, since the scalar Lagrangian has at least two derivatives.
Thus we learn that $\epsilon, \delta$ can at most be of order 
\be
\label{epsdelo}
\epsilon,\delta \sim \bigg({ {\dot{\bar{\phi}}} \over H}\bigg)^2,
\ee
and   eq.(\ref{condbsl})  then leads to eq.(\ref{conddd}).
The equations, (\ref{condslra}),  (\ref{condbsl}) and (\ref{conddd}) are what we will use in our derivation of the Ward identities. 

We end with a few comments which are  of relevance for the discussion in section \ref{solvewardid}, where we estimate the normalization 
 of the homogeneous term $S_h$ in the solution of  the Ward identities.  We begin by noting that
when the higher derivative terms are important for the metric, $H^2$ will not be given in terms of $V$ by eq.(\ref{hubble}). Instead, the relation will be more complicated and have the form
\be
\label{newhubble}
H^2\, f \Big({H\over  \Lambda}\Big)= V,
\ee
where $f$ is a function which depends on the higher derivative contributions. 
Now as long as the function $f \sim O(1)$, we  get 
\be
\label{approxh}
H^2 \sim V.
\ee 
Taking a time derivative then gives, 
\be
\label{tdera}
{\dot{H}\over H^2} \sim {V'\over H^2} \, {\dot{\bar{\phi}}\over H}\,.
\ee
Using eq.(\ref{epsdelo}) then leads to 
\be
\label{sceqa}
\dot{\bar{\phi}} \sim {V'\over H}.
\ee

It follows from eq.(\ref{approxh}) and eq.(\ref{sceqa}) that the general slow roll case is in fact quite analogous to the canonical slow roll model. 
In particular, it follows from eq.(\ref{approxh}), eq.(\ref{sceqa}) that 
\be
\label{approxrel}
{\dot{\bar{\phi}}\over H}\sim \sqrt{\epsilon}\,,
\ee 
and also that in the slow roll expansion in general, an extra time derivative leads to a suppression by a factor of $\epsilon$. 

The function $f$ in eq.(\ref{newhubble}) has the limiting behaviour $f\rightarrow 3$ when $ {H\over  \Lambda} \rightarrow 0$.
Eq. (\ref{approxh}) is therefore a reasonable assumption if $f\sim O(1)$  also for ${H\over \Lambda} \sim O(1)$, but it could be a bad approximation if $f$ becomes big for ${H \over \Lambda} \sim O(1)$.


\section{The Ward Identities}
\label{wardderiv}
We now turn to a discussion of the Ward identities. 
It is convenient to first consider the case of pure de Sitter space, with no corrections, and then consider the inflationary spacetime. 
\subsection{de Sitter Space}

In de Sitter space the metric perturbations $\gamma_{ij}$ and the scalar 
perturbation $\delta \phi$ both freeze out and become time independent 
at sufficiently late time, when their physical spatial momenta ${|\bf k|\over a}$
become much smaller than $H$. 

The late time wave function is then a functional of these variables, as discussed in eq.\eqref{wf1}. 
As was mentioned above in the comments after eq.\eqref{wf2}, our choice eq.\eqref{gaugec} does not fix the gauge completely.
In the discussion below, it will be sometimes convenient to fix the remaining time reparametrization freedom, eq.\eqref{timere}, by setting the late time value of $\zeta$ to vanish,\footnote{This choice will be referred to as gauge A in section \ref{chgauge}.}
\be
\label{gaugea}
\zeta =0.
\ee
It is possible to do this for a suitable choice of $\epsilon(\vec[x])$ because at late times, when $v^i$ in eq.\eqref{videff} vanishes, $\zeta$ transforms under
\begin{equation*}
t \rightarrow t + \epsilon(\vec[x])
\end{equation*}
as
\begin{equation*}
\zeta \rightarrow \zeta - H \epsilon(\vec[x]).
\end{equation*}

After this additional gauge fixing eq.\eqref{gaugea}, the Ward identities of special conformal transformations are then derived in this
gauge  by considering  a combined spatial 
reparametrization and time reparametrization,
\begin{align}
x^i   \rightarrow   x^i -2 (b_j x^j) x^i \, + \, & b^i \bigg(\sum_j (x^j)^2
- {e^{-2Ht}\over H^2}\bigg) \label{spcon2}, \\
t  \rightarrow   t \, + \, &2\,{b_j x^j \over H} \label{tspcon2}, 
\end{align} 
which preserve the gauge condition eq.(\ref{gaugea}). Before proceeding, let us note that
the special conformal transformations are specified by three parameters, $b^i, i=1, \cdots 3$. 
Also, note that the last term in eq.\eqref{spcon2}, which goes like 
$b^i {e^{-2Ht}\over H^2}$, can be dropped at late time. 

The invariance of the wave function under the combined transformation, eq.(\ref{spcon2}), eq.(\ref{tspcon2}),  gives rise to
constraints on the coefficient functions in eq.(\ref{wf1}). In particular, for the coefficient function $\langle OOO \rangle$ in eq.(\ref{wf1}),
which is the coefficient of the term cubic in $\delta \phi$ in the wave function,
this leads to the condition,
\be
\label{delthrm}
{\cal L}^{\vec[b]}_{\vecs[k,1]} \langle O(\vecs[k,1]) O(\vecs[k,2]) O(\vecs[k,3]) \rangle' +
{\cal L}^{\vec[b]}_{\vecs[k,2]} \langle O(\vecs[k,1]) O(\vecs[k,2]) O(\vecs[k,3]) \rangle'+
{\cal L}^{\vec[b]}_{\vecs[k,3]} \langle O(\vecs[k,1]) O(\vecs[k,2]) O(\vecs[k,3]) \rangle'= 0,
\ee
where ${\cal L}^{\vec[b]}_{\vec[k]}$ is the differential operator
\be
\label{delom}
{\cal L}^{\vec[b]}_{\vec[k]} = 2 \, \Big(\vec[k] \cdot \frac{\del}{\del \vec[k]}\Big) \, \Big(\vec[b] \cdot 
\frac{\del}{\del \vec[k]}\Big) - (\vec[b] \cdot \vec[k]) \, \Big(\frac{\del}{\del \vec[k]} \cdot \frac{\del}{\del \vec[k]}\Big).
\ee
The prime symbols on the correlation functions in eq.(\ref{delthrm}) denote the 
correlation functions with the momentum conserving delta function stripped off:
\be
\label{defprime}
\langle O(\vecs[k,1]) O(\vecs[k,2]) O(\vecs[k,3]) \rangle = (2 \pi)^3 \, \delta^3(\vecs[k,1]+\vecs[k,2]+\vecs[k,3]) \, \langle O(\vecs[k,1]) O(\vecs[k,2]) O(\vecs[k,3]) \rangle'.
\ee
We will follow a similar convention in this paper  for other correlation functions as well. 

It is worth giving some more details leading to  eq.(\ref{delthrm}). Since the asymptotic 
value of $\delta \phi$ is time independent, it only transforms under the 
spatial reparametrization, eq.\eqref{spcon2}, 
\begin{align}
\begin{split}
\label{transdp}
\delta \phi & \rightarrow   \delta \phi + \delta(\delta\phi(\vec[x])),\\
\delta (\delta\phi(\vec[x]))  & = \left( 2(\vec[b]\cdot\vec[x]) x^{i} - 
\vec[x]^2 b^{i} \right) \partial_i (\delta \phi(\vec[x])).
\end{split}
\end{align}
Requiring that the wave function is invariant gives rise to the condition
\be
\label{invw2}
\Psi[\delta \phi]= \Psi[\delta \phi + \delta (\delta \phi)].
\ee
For the coefficient $\langle OOO \rangle$ in position space this leads to the relation, 
\be
\label{delthr}
\langle (\delta O(\vec[x])) O(\vec[y]) O(\vec[z]) \rangle + \langle O(\vec[x])
(\delta O(\vec[y])) O(\vec[z]) \rangle  + \langle O(\vec[x]) O(\vec[y]) (\delta O(\vec[z])) \rangle = 0,
\ee
where,
\be
\label{delo}
\delta O(\vec[x]) = \left( \vec[x]^2 b^{i} - 2(\vec[b]\cdot\vec[x]) x^{i} \right) \partial_i O(\vec[x]) - 6(\vec[b]\cdot\vec[x]) O(\vec[x]).
\ee
Eq.(\ref{delo})  becomes eq.\eqref{delthrm} in momentum space. 
The wave function also depends on $\gamma_{ij}$, which
transforms under eq.(\ref{spcon2}), eq.(\ref{tspcon2}), but the resulting terms 
are not relevant for obtaining the identity eq.(\ref{delthr}) and we omit them here.

The Ward identity for scale transformations can be derived in a similar way by requiring
the invariance of the wave function
under the coordinate transformation 
\be
\label{scaltran}
t \rightarrow t + \lambda, \, \, x^i \rightarrow e^{-H\lambda} \, x^i \approx (1 - H\lambda) x^i.
\ee
The scalar perturbation $\delta \phi$ transforms under this as
\begin{align}
\begin{split}
\label{delphsch}
\delta\phi \rightarrow \delta\phi \, + \, \delta(\delta\phi), \\
\delta(\delta\phi) = H\lambda \, x^i\del_i\delta\phi.
\end{split}
\end{align}
For the coefficient function $\langle OOO \rangle$ this gives the relation
\be
\label{scalwardh}
\langle (\delta O(\vec[x])) O(\vec[y]) O(\vec[z]) \rangle + \langle O(\vec[x]) 
(\delta O(\vec[y])) O(\vec[z]) \rangle  + \langle O(\vec[x]) O(\vec[y]) 
(\delta O(\vec[z])) \rangle = 0\,,
\ee
where $\delta O(\vec[x])$ is now given by
\be
\label{deloscal}
\delta O(\vec[x]) = H\lambda \left(3+x^i \del_i\right) O(\vec[x]).
\ee
The first term on the RHS of eq.\eqref{deloscal} arises as follows. Each factor of $\delta \phi(\vec[x])$ in the cubic 
term in the wave function, eq.(\ref{wf1}), is accompanied by an integration measure,
$\int d^3 x \sqrt{g(\vec[x])}$. Since we are in the gauge $\zeta=0$,
$\sqrt{g}=1$ and does not change under the transformation eq.(\ref{scaltran}).
The change in the measure $d^3x$ under eq.(\ref{scaltran}) then gives rise to this first term. 
We note that eq.\eqref{scalwardh} is what we would expect for an operator of dimension $3$ in a CFT. 
In momentum space eq.(\ref{scalwardh}) becomes
\be
\label{momwardsc}
\bigg( \sum_{a=1}^3 \vec[k_a] \cdot \frac{\partial}{\partial{\vec[k_a]}} \bigg) \langle O(\vecs[k,1])
O(\vecs[k,2]) O(\vecs[k,3])\rangle = 0.
\ee

\subsection{Inflationary Spacetime}

Now let us consider departures from the conformally invariant case which arise during inflation. 
In general, the metric begins to differ from the de Sitter case and this in turn affects the asymptotic
behavior of the various perturbations. 
It turns out that for the limited purpose of deriving the Ward identities of interest, the departures 
of the metric from de Sitter space can be neglected. This is because these departures, which arise 
because $H$ is no longer a constant, are proportional to $\epsilon, \delta$, eq.\eqref{defslowroll},
whereas the Ward identity we  seek arises at order ${\dot{\bar{\phi}}\over H}$. Since we have argued 
 that the condition eq.\eqref{conddd}, which is true in the canonical slow roll theory is 
also true more generally,
it is consistent to take  the background metric to be de Sitter space while keeping corrections of order
	${\dot{\bar{\phi}} \over H}$. 

This  approximation leads to  considerable simplification.  The asymptotic behavior of perturbations continues to
be that of de Sitter space. As a result, it is quite straightforward to connect with the analysis above in de Sitter space. 

\subsubsection{Choice of Gauge}
\label{chgauge}
There is one subtlety in the inflationary case which needs to be kept in mind though. A variable which is often used to describe scalar perturbations in inflation 
is the variable ${\cal R}$, given by 
\be
\label{defMS}
{\cal R} = \zeta - {H \over {\dot{\bar{\phi}} } } \, \delta\phi\,.
\ee
The  variable ${\cal R}$  has the advantage that it is  invariant under linearized coordinate transformations, and is also constant outside the horizon. However, since ${\dot{\bar\phi}}$ appears in the denominator on the RHS, taking 
 the $\dot{\bar{\phi}}\rightarrow 0$ limit, when the de Sitter description should become a good one, can sometimes be confusing when working  directly in terms of ${\cal R}$.

The simplest way to deal with this complication is to use two different gauges.  While the perturbations are inside the horizon and evolving, one can  work in the gauge where eq.(\ref{gaugea})  is true. We refer to this as gauge A below. In this gauge the scalar perturbation is given by $\delta \phi$ which  behaves in a smooth way, with a  well defined Lagrangian for example,  in the de Sitter limit. Once the perturbations leave the horizon, one can then go over to the gauge where
\be
\label{gaugeb}
\delta \phi=0
\ee
is true. In this gauge the scalar  perturbation is given by $\zeta$  and is a constant outside the horizon, so that the correlation functions in terms of $\zeta$  are time independent. We call this gauge B below.  The required coordinate transformation is a time reparametrization eq.(\ref{timere}), with a suitably chosen time independent parameter $\epsilon({\bf x})$. At the linearized level the variable $\zeta$ in gauge B is related to the variable $\delta \phi$ in gauge A by
\be
\label{relzpa}
\zeta = - \, {H \over \dot{\bar{\phi}}} \, \delta \phi\,.
\ee
Having calculated the correlation functions  in gauge A it is a straightforward exercise, only  involving a  change of variables, to go over to gauge B.

This is in fact the procedure we will follow below. To begin, we will work in gauge A and construct the wave function in terms of $\delta \phi$ and the remaining degrees of freedom in the metric $\gamma_{ij}$. We can think of this wave function as being constructed in the epoch when the perturbations of interest are exiting the horizon. It  will take the form given in eq.(\ref{wf1}). We will then obtain relations between various coefficient functions of this wave function by demanding that it is invariant under suitable time and spatial reparametrizations.  Then we will change the gauge and go to gauge B, and  recast these relations now between correlation functions of $\zeta$,  which are conserved outside the horizon. 

One more comment is in order before we proceed. Although the traceless component of the metric perturbation, ${\widehat\gamma}_{ij}$, eq.(\ref{defzeta}), will not play much of a  role in the following discussion, we have in mind carrying out a spatial reparametrization eq.(\ref{spacere}) so that at late time ${\widehat \gamma}_{ij}$ satisfies the condition, 
\be
\label{gaugecondb}
\partial_i{\widehat \gamma}^{ij}=0.
\ee
Indeed, only after this gauge fixing is ${\cal R}$ given by eq.(\ref{defMS}). 

\subsubsection{The Ward Identities}
 
Setting $\zeta=0$, eq.(\ref{gaugea}), to derive the Ward identity of special conformal transformations, we again choose the spatial  and time reparametrizations,
eq.(\ref{spcon2}), eq.(\ref{tspcon2}), and demand that the wave function is invariant under them.
The only new change is that since we are also keeping effects of order ${\dot{\bar{\phi}}}\over H$ 
now, the change in the scalar perturbation $\delta \phi$ has an extra term compared to eq.\eqref{transdp}.

This extra term arises as follows. One wants the full inflaton field, eq.\eqref{infla}, to transform 
like a scalar under the coordinate transformation eq.\eqref{spcon2}, \eqref{tspcon2}. That is, 
denoting a generic coordinate transformation as 
\be
\label{genctr}
x^\mu \rightarrow x^\mu + \epsilon^\mu ( x),
\ee
(where $\mu = 0,1,2,3$), 
$\phi $ should transform as 
\be
\label{transp2}
\phi \rightarrow \phi - \epsilon^\mu \partial_\mu \phi.
\ee
It is easy to see that this gives rise to an extra term in the transformation for $\delta \phi$, 
so that, to this order
\be
\label{trans33}
\delta \phi \rightarrow \delta \phi + \delta (\delta \phi) + \tilde{\delta}(\delta \phi),
\ee
where $\delta(\delta \phi)$ is the same as in eq.\eqref{transdp} and $\tilde{\delta}(\delta\phi)$, 
the extra contribution, is 
given by
\be
\tilde{\delta}(\delta \phi (\vec[x])) = - \, 2(\vec[b]\cdot\vec[x]) \frac{\dot{\bar{\phi}}}{H}.
\ee

Now demanding that the wave function is invariant under the full change of $\delta \phi$ gives 
rise to a modified Ward identity, which takes the form

\begin{align}
\label{wardidsct}
\begin{split}
{\cal L}^{\vec[b]}_{\vecs[k,1]} \langle O(\vecs[k,1]) O(\vecs[k,2]) O(\vecs[k,3]) \rangle' & +
{\cal L}^{\vec[b]}_{\vecs[k,2]} \langle O(\vecs[k,1]) O(\vecs[k,2]) O(\vecs[k,3]) \rangle'+
{\cal L}^{\vec[b]}_{\vecs[k,3]} \langle O(\vecs[k,1]) O(\vecs[k,2]) O(\vecs[k,3]) \rangle' \\ 
& = 2 \, \frac{\dot{\bar{\phi}}}{H} \bigg[ \vec[b] \cdot \frac{\del}{\del \vecs[k,4]} \bigg] 
\bigg\{ \langle O(\vecs[k,1]) O(\vecs[k,2]) O(\vecs[k,3]) O(\vecs[k,4]) \rangle' \bigg\rvert_{\vecs[k,4] 
\rightarrow 0} \bigg\},
\end{split}
\end{align}
where ${\cal L}^{\vec[b]}_{\vec[k]}$ is the same as defined in eq.\eqref{delom}.

Similarly, for the scaling transformation, eq.\eqref{scaltran}, we get the Ward identity
\begin{equation}
\label{wardidscal}
\bigg( \sum_{a=1}^3 \vec[k_a] \cdot \frac{\partial}{\partial{\vec[k_a]}} \bigg) \langle O(\vecs[k,1])
O(\vecs[k,2]) O(\vecs[k,3])\rangle  = \, \frac{\dot{\bar{\phi}}}{H} \, 
\langle O(\vecs[k,1]) O(\vecs[k,2]) O(\vecs[k,3]) O(\vecs[k,4]) \rangle
\bigg\rvert_{\vecs[k,4] \rightarrow 0}.
\end{equation}
Eq.\eqref{wardidscal} and especially eq.\eqref{wardidsct} are some of the main results of this paper.

So far our discussion was in terms of the coefficient functions which appear in the wave function. 
It is useful to express the results in terms of correlation functions of perturbations.  
The expectation values of correlators involving $\delta \phi$ can be obtained from the wave
function in the standard fashion. For example, the two point function is
\be
\label{sctwopt}
\langle \delta \phi(\vec[x]) \delta \phi(\vec[y]) \rangle ={\int [{\cal D} \delta \phi] [{\cal D} \gamma_{ij}] \; |\Psi|^2  \; \delta \phi (\vec[x]) \,
\delta \phi (\vec[y]) \over \int [{\cal D}\delta \phi] [{\cal D}\gamma_{ij}] \; |\Psi|^2}.
\ee
From eq.(\ref{wf1}) we see that in momentum space this gives, 
\begin{align}
\label{sctwoptb}
\langle \delta \phi(\vec[k]) \delta \phi(\vec[k']) \rangle &= (2\pi)^3 \delta^3(\vec[k]+\vec[k']) \, \half \frac{H^2}{M_{Pl}^2} \, \frac{1}{\langle O(\vec[k]) O(\vec[k']) \rangle'} \\ &= (2\pi)^3 \delta^3(\vec[k]+\vec[k']) \, \frac{H^2}{M_{Pl}^2} \frac{1}{2k^3} \, ,\label{sctwoptc}
\end{align}
where we have used eq.(\ref{defoo}). 

Although it will not be very relevant for the present discussion, let us note that the RHS
of eq.(\ref{sctwopt}) is slightly imprecise. To make the sum over metrics well defined, 
the remaining gauge redundancy must also be removed. This is a general feature when calculating
expectation values, \cite{Ghosh:2014kba}. While we are not being very explicit about this, we always have in 
mind fixing this redundancy by also taking ${\hat \gamma}_{ij}$ to be transverse, eq.(\ref{gaugecondb}).
Note that   $\zeta$ is already set to vanish in the gauge we are working with so far, eq.(\ref{gaugea}).

Once the correlation functions for $\delta \phi$ have been obtained, we can change gauge and go over to gauge B, eq.(\ref{gaugeb}), as was discussed in subsection \ref{chgauge} above. 

For the two point function, we see from eq.(\ref{sctwoptc}), eq.(\ref{gaugea}) and eq.(\ref{defMS}) that the variable ${\cal R}$ has the two point function, 
\be
\label{twoptms}
\langle {\cal R}(\vec[k]) {\cal R}(\vec[k'])\rangle = (2\pi)^3 \delta^3(\vec[k]+\vec[k']) \, \frac{H^2}{M_{Pl}^2} \, \frac{H^2}{{\dot{\bar{\phi}}}^2} \, \frac{1}{2k^3},
\ee
which is the standard result. 
In gauge B where eq.(\ref{gaugeb}) is met, 
\be
\label{relrz}
{\cal R}=\zeta \,.
\ee
Thus, eq.(\ref{twoptms}) leads to, 
\be
\label{twoptzeta}
\langle \zeta (\vec[k]) \zeta (\vec[k'])\rangle = (2\pi)^3 \delta^3(\vec[k]+\vec[k']) \, \frac{H^2}{M_{Pl}^2} \, \frac{H^2}{{\dot{\bar{\phi}}}^2} \, \frac{1}{2k^3}.
\ee
For completeness, we also note that the graviton two-point function is given by 
\be
\label{twoptgamma}
\langle \gamma_s(\vecs[k,1]) \gamma_{s'}(\vecs[k,2]) \rangle = (2\pi)^3 \delta^3(\vecs[k,1]+\vecs[k,2]) \, \delta_{s,s'} \, \frac{H^2}{M_{Pl}^2} \, \frac{1}{k_1^{\,3}} ,
\ee
where $\gamma_s = \half \, \gamma_{ij} \, \epsilon_s^{\,ij}$.

At linear order the variable  $\zeta$ in gauge B is related to $\delta \phi$ in  gauge A  by eq.(\ref{relzpa}).
When we consider the three point function things get a little more complicated in going over to gauge B. Since the three point function is suppressed 
(due to the factor of $\dot{\bar {\phi}}$ on the RHS of eq.(\ref{wardidsct})) the relation, eq.(\ref{relzpa}),
is needed to second order. It turns out to be \footnote{It follows from inverting eq.(D.8) in \cite{Ghosh:2014kba} to obtain $\zeta$ in terms of $\delta \phi$.}
\be
\label{relzpb}
\zeta = - \, {H \over {{\dot{\bar{\phi}}}}} \, \delta \phi + {1\over2} {H \over {{\dot{\bar{\phi}}}}} \bigg({\dot{H} \over H{{\dot{\bar{\phi}}}}} - \frac{\ddot{\bar{\phi}}}{{\dot{\bar{\phi}}}^2} \bigg) \delta\phi^{\,2}.
\ee
It was shown in \cite{Maldacena:2002vr} that $\zeta$ in gauge B is in fact constant outside the 
horizon, and since we have gauge fixed completely, it is also a physical observable. 
This makes it a  convenient variable to use.  
From eq.(\ref{relzpb}) and eq.\eqref{ooophi3} we get that \footnote{Note that the second term on the RHS of eq.\eqref{tpzeta} is of the same order as the first term, $\langle O O O\rangle'$. For instance, $\frac{\ddot{\bar{\phi}}}{{\dot{\bar{\phi}}}^2} = \left( \frac{\ddot{\bar{\phi}}}{H \dot{\bar{\phi}}}\right) \left(\frac{H}{\dot{\bar{\phi}}}\right) = \frac{\delta}{\sqrt{2\epsilon}} \approx \sqrt{\epsilon}$.}
\begin{align}
\label{tpzeta}
\begin{split}
\langle \zeta(\vec[k_1]) \zeta(\vec[k_2]) \zeta(\vec[k_3]) \rangle = & {1\over 4} \, {H^4\over M_{pl}^4} \, {H^3\over {\dot{\bar{\phi}}^3}} \, (2 \pi)^3 \delta^3(\vec[k_1]+\vec[k_2]+\vec[k_3]) \, {1\over \prod_{a=1}^3 k_a^{\,3}}  \\ & \bigg[-\langle O(\vec[k_1]) O(\vec[k_2]) O(\vec[k_3]) \rangle'
+\bigg({\dot{H} \over H{{\dot{\bar{\phi}}}}} - {\ddot{\bar{\phi}} \over {\dot{\bar{\phi}}}^{\,2}} \bigg) \Big(\sum_{a=1}^3 k_a^{\,3}\Big)\bigg].
\end{split}
\end{align}

Similarly, the four point function to leading order is given by 
\be
\label{fpzeta}
\begin{split}
\langle \zeta(\vecs[k,1]) \zeta(\vecs[k,2]) \zeta(\vecs[k,3]) \zeta(\vecs[k,4]) \rangle = \, &\langle \zeta(\vecs[k,1]) \zeta(\vecs[k,2]) \zeta(\vecs[k,3]) \zeta(\vecs[k,4]) \rangle_{CF} \\ & + \langle \zeta(\vecs[k,1]) \zeta(\vecs[k,2]) \zeta(\vecs[k,3]) \zeta(\vecs[k,4]) \rangle_{ET}\,.
\end{split}
\ee
The two terms on the RHS of eq.(\ref{fpzeta}) were calculated in \cite{Seery:2008ax} and  \cite{Ghosh:2014kba}, and are also given in eq.(\ref{fpzetacf}) and eq.(\ref{fpzetaet}) of appendix \ref{appsuboooo}. In particular, $\langle \zeta\zeta\zeta\zeta \rangle_{ET}$ is determined in terms of the $\langle O O T_{ij} \rangle$ correlator, and therefore completely fixed by conformal invariance, see \cite{Mata:2012bx}.

By inverting eq.(\ref{tpzeta}) and eq.(\ref{fpzeta}), one can express $\langle OOO \rangle$ and $\langle OOOO \rangle$ 
in terms of the three point $\zeta$ correlator $\langle \zeta \zeta \zeta \rangle$, and $\langle \zeta\zeta\zeta\zeta \rangle_{CF}$ respectively, eq.\eqref{fpzetacf}. 
It turns out that the contribution of $\langle \zeta \zeta\zeta\zeta \rangle_{ET}$ to the RHS of the Ward identities vanishes. As a result, eq.\eqref{wardidsct} and \eqref{wardidscal} then become
\begin{align}
\begin{split}
\label{wardsctzt}
{\widehat {\cal L}}^{\vec[b]}_{\vecs[k,1]} \langle \zeta(\vecs[k,1]) \zeta(\vecs[k,2]) &\zeta(\vecs[k,3]) \rangle' + {\widehat{\cal L}}^{\vec[b]}_{\vecs[k,2]} \langle \zeta(\vecs[k,1]) \zeta(\vecs[k,2]) \zeta(\vecs[k,3]) \rangle'+
{\widehat{\cal L}}^{\vec[b]}_{\vecs[k,3]} \langle \zeta(\vecs[k,1]) \zeta(\vecs[k,2]) \zeta(\vecs[k,3]) \rangle' \\ 
& = -\, 4 \, \frac{M_{Pl}^2}{H^2} \, \frac{\dot{\bar{\phi}}^{\,2}}{H^2} \, \bigg[ \vec[b] \cdot \frac{\del}{\del \vecs[k,4]} \bigg] \bigg\{ k_4^{\,3}~ \langle \zeta(\vecs[k,1]) \zeta(\vecs[k,2]) \zeta(\vecs[k,3]) \zeta(\vecs[k,4]) \rangle' \bigg\rvert_{\vecs[k,4] \rightarrow 0}\bigg\},
\end{split}
\end{align}
and
\begin{equation}
\label{wardscalzt}
\bigg[6+\sum_{a=1}^3 \vec[k_a] \cdot \frac{\partial}{\partial{\vec[k_a]}} \bigg] \langle \zeta(\vecs[k,1]) \zeta(\vecs[k,2]) \zeta(\vecs[k,3])\rangle' = - \, 2 \, \frac{M_{Pl}^2}{H^2} \, \frac{\dot{\bar{\phi}}^2}{H^2} \, k_4^{\,3} \, \langle \zeta(\vecs[k,1]) \zeta(\vecs[k,2]) \zeta(\vecs[k,3]) \zeta(\vecs[k,4]) \rangle' \bigg\rvert_{\vecs[k,4] \rightarrow 0},
\end{equation}
with \footnote{We remind the reader that a prime symbol on a correlator denotes that the momentum conserving delta function has been removed, see eq.(\ref{defprime}).}
\be
{\widehat{\cal L}}^{\vec[b]}_{\vec[k]} = {\cal L}^{\vec[b]}_{\vec[k]}+6\,\left[ \vec[b] \cdot \frac{\del}{\del \vec[k]}\right],
\ee
and ${\cal L}^{\vec[b]}_{\vec[k]}$ as given in eq.(\ref{delom}).

In this way, we see that the 
Ward identities eq.\eqref{wardidsct} and eq.\eqref{wardidscal} derived above impose conditions on the physically observable three 
and four point correlators. Some of these Ward identities have been discussed  in the literature before, e.g.,
setting ${\vec[b]} \propto {\vecs[k,4]}$ in eq.(\ref{wardsctzt}) gives eq.(37) in \cite{Creminelli:2012ed}.

 
\section{Comments on the Ward Identities}
\label{commentward}
Here we comment on the Ward identities obtained above in more detail. 

\subsection{The Canonical Slow Roll Model as a Check}
 The Ward identities obtained above  can be checked in the canonical slow roll model, 
eq.\eqref{actslroll}, and shown to hold.  For the slow roll model eq.\eqref{actslroll}, the three point function was 
obtained in \cite{Maldacena:2002vr}. 
The corresponding cubic coefficient function can be easily calculated, as discussed in 
appendix \ref{appsubooo}, and is given by 
\be
\label{cusl}
\langle O(\vecs[k,1]) O(\vecs[k,2]) O(\vecs[k,3]) \rangle' = - \, {3\epsilon + 4\delta \over {2 \sqrt{2\epsilon}}} \sum_{a} k_{a}^{\,3} - \half \sqrt{2\epsilon} \, \bigg( \half \sum_{a \neq b} k_a k_b^2 + \frac{4}{k_t} \sum_{a>b} k_a^2 k_b^2 \bigg),
\ee
where $k_a \equiv |\vecs[k,a]|$, and $k_t = k_1 + k_2 + k_3$.

The four point function in this model was discussed in \cite{Seery:2008ax} and also in \cite{Ghosh:2014kba}. The corresponding 
coefficient function is given in eq.(4.33) of \cite{Ghosh:2014kba} (see appendix \ref{appsuboooo} of this paper). 

To check the Ward identity for scale invariance eq.(\ref{wardidscal}), we note that 
since $\langle OOO\rangle '$ in eq.(\ref{cusl}) is cubic in momenta, the LHS of eq.(\ref{wardidscal})
vanishes.  From eq.(6.21) and (6.22) of \cite{Ghosh:2014kba}, it is easy to check that the RHS of eq.(\ref{wardidscal})
also vanishes when $\vecs[k,4] \rightarrow 0$. Thus the Ward identity eq.\eqref{wardidscal} holds. 

The check for the Ward identity of special conformal transformations, eq.(\ref{wardidsct}),  is more 
complicated because the four point coefficient function $\langle OOOO\rangle$  is an unwieldy large 
expression. Nevertheless, using Mathematica one can check that it is indeed valid. It is easy to see that the function 
$k^{\,3}$ satisfies the condition,
\be
\label{annk}
{\cal L}^{\vec[b]}_{\vec[k]} \, (k^{\,3}) =0,
\ee
where the operator ${\cal L}^{\vec[b]}_{\vec[k]}$ is defined in eq.(\ref{delom}). 
The non-trivial contribution for the LHS of the Ward identity eq.\eqref{wardidsct} comes therefore from the second term in eq.(\ref{cusl}).
The $\langle OOOO \rangle$   coefficient function has two kinds of contributions, denoted by $\widehat{W}^S$ and $\widehat{R}^S$ (see eq.(\ref{fourptwoapp})). Of these, only the $\widehat{R}^S$ term contributes.

\subsection{Constraint on the Magnitude of the Three Point Function}

We see from eq.\eqref{cusl} that the cubic coefficient function $\langle OOO\rangle $ vanishes in the canonical slow roll model in the limit when 
the slow roll parameters  vanish. This is well known
 and is  responsible for the small magnitude of the non-Gaussianity in this model.
One can argue more generally   that the cubic coefficient $\langle OOO \rangle$ must vanish in the limit when all the slow roll parameters vanish. In the gravity calculation, this happens because  in this limit  $\delta \phi$ becomes a massless scalar field in de Sitter space with no potential, and therefore does not have a three point function. From the point of view of conformal invariance and the related CFT, in this limit the corresponding operator $O$ is exactly marginal, and in a CFT it is well known that the three point function of an exactly marginal operator vanishes. This is analogous to what happens in $2$ dimensional CFT, see for example section (15.8) of \cite{Polchinski}. If this three point function would not vanish then $\langle O\rangle $ for example would have a log divergence at second order in perturbation theory, leading to a non-zero beta function for $O$. Thus, on general grounds, we know that the expectation value for the scalar three point function should be suppressed. 

The Ward identity, eq.(\ref{wardidsct}),  allows us to estimate the magnitude of the three point function once non-vanishing values for the slow roll parameters are taken into account. Since the quartic coefficient function $\langle OOOO\rangle $ is not expected to vanish in the de Sitter limit, we see 
from eq.(\ref{wardidsct})  that the RHS is of order ${\dot {\bar{\phi}}\over H}$.  From this, it follows  quite naturally  that the $\langle OOO\rangle$  coefficient function will  be of order ${\dot{\bar{\phi}} \over H}$. So we see  that as long as conformal invariance is an approximate symmetry, the three point scalar correlator  will be of order its value in the canonical slow roll model, eq.(\ref{tpzeta}), and therefore be  small. Although the functional form is not the same as in the  standard $f_{NL}$ parametrization, to get a rough idea,  this magnitude corresponds to an  $f_{NL}\sim O\big(({\dot{\bar{\phi}}\over H})^2\big)$.   If observationally a scalar non-Gaussianity is observed in the near future,  its magnitude would most likely be much bigger. Thus 
the considerations of this 
paper show that such an observation would not only rule out the canonical slow roll model, but more generally any model which preserves approximate conformal invariance during inflation. 
Note that in our conventions, the scalar and tensor two point correlators are given in eq.(\ref{twoptms}) and eq.(\ref{twoptgamma}). 

There is one important caveat to the above statement. As will be discussed in the next subsection, the Ward identity eq.(\ref{wardidsct}) does not uniquely determine the coefficient function $\langle OOO\rangle $ and thus the scalar three point function $\langle \zeta \zeta \zeta \rangle$, in terms of $\langle OOOO\rangle $. The remaining freedom corresponds to the three point function of a dimension 3 primary scalar operator in a CFT, $S_h$,  with an arbitrary overall normalization. However, as we argue there, with generic assumptions,  in the slow roll approximation this
 normalization is expected  to be  small, making any such contribution to $\langle OOO\rangle $ even more suppressed than that which originates from the $\langle OOOO\rangle $ source term. In case these generic assumptions are  somehow not met, and the normalization is bigger making $S_h$ dominate,  the functional form of the three point function will be fixed (upto a contact term) and this  possibility can therefore  also be checked observationally.

\subsection{Solving the Ward Identities to Determine the  Three Point Function}
\label{solvewardid}
In this  subsection, we investigate the question of uniqueness: given a four point coefficient function $\langle OOOO\rangle$, to what extent do the Ward identities, eq.(\ref{wardidsct}) and eq.(\ref{wardidscal}), fix the three point coefficient  function,   $\langle OOO\rangle $. We find, not surprisingly,  that there is very little freedom that remains. It corresponds to adding to the three point coefficient function a term whose form is the same as  the three point function of a dimension $3$ operator in a CFT, $S_h$. 
The momentum dependence of  this additional function is completely fixed, and all that is left undetermined is its overall normalization\footnote{There is also an additional constant associated with a contact term, see below.}.
  Besides this normalization our conclusion is therefore that the three point function is completely fixed in terms of the four point function. This is an interesting result  because unlike the three point function, the four point function, $\langle OOOO\rangle $, does not vanish in the conformally invariant case. By relating the two, we learn that the freedom allowed by the approximate conformal symmetry for the three point function is about the same as
 that in the four point function. Towards the end of this section we argue that the normalization constant for the additional term $S_h$ should be suppressed generically in  the slow roll approximation, so that even this remaining ambiguity is not important.

The Ward identities are   in the form of  linear differential equations for $\langle OOO\rangle '$, with $\langle OOOO\rangle '$ appearing on the RHS as a source or  inhomogeneous term. Suppose there are two solutions for $\langle OOO\rangle '$ allowed by eq.(\ref{wardidsct}), eq.(\ref{wardidscal}).  Let us denote their  difference as
\be
\label{defSh}
\langle OOO\rangle' _1-\langle OOO\rangle' _2= S_h(\vecs[k,1], \vecs[k,2], \vecs[k,3]).
\ee
It is clear that $S_h$ solves the homogeneous equations, 
\be
\label{homeqscale}
\Big( \sum_{a=1}^3 \vecs[k,a] \cdot \frac{\del}{\del \vecs[k,a]} \Big) \, S_h(\vecs[k,1], \vecs[k,2], \vecs[k,3])= 3 \, S_h(\vecs[k,1], \vecs[k,2], \vecs[k,3])
\ee
and 
\be
\label{homeqsct}
\big( \sum_{a=1}^3 {\cal L}^{\vec[b]}_{\vecs[k,a]} \big) \, S_h(\vecs[k,1], \vecs[k,2], \vecs[k,3])=0.
\ee
The RHS in eq.(\ref{homeqscale}) arises because the delta function has been removed in defining $\langle OOO\rangle '$. 
By comparing with eq.(\ref{delthrm}) and eq.(\ref{momwardsc}),  we see that these are exactly the equations satisfied by the three point function of a dimension $3$ operator in the CFT.  

It is well known that the three point function of a dimension $3$ primary in a CFT is fixed in position space upto overall normalization. 
We find a similar result on analyzing the two equations  eq.(\ref{homeqscale}) and eq.(\ref{homeqsct}) in momentum space.
Upto an additional constant, which affects only contact terms in position space, the only freedom in $S_h$ allowed is the overall normalization. 
Details of this analysis are given in the appendix \ref{appSh}.

Since $\langle OOO\rangle $ conserves overall momentum, it is easy to see that $S_h$ can be taken to be a function of only the three scalars, $k_a, \,a=1,\cdots 3$. 
Our analysis in appendix \ref{appSh}  then gives,
\be
\label{shfinal}
S_h(k_1, k_2, k_3)  =  N \, \frac{1}{3}  \Big[ \ln(\lambda) \Big(\sum_{a=1}^3 k_a^{\,3}\Big) + \ln(\sum_{a=1}^3 k_a)  \Big(\sum_{b=1}^3 k_b^{\,3}\Big) - \sum_{a \neq b} k_a k_b^2 + k_1 k_2 k_3 \Big],
\ee
where $\lambda$ is a short distance cut-off which is introduced in obtaining the solution. As discussed in appendix \ref{appSh}, in obtaining this final form for the solution we have also imposed conditions which arise from the operator product expansion. $N$ is the overall undetermined normalization, and $\ln(\lambda)$ is the extra coefficient which multiplies the contact term $(\sum_a  k_a^3$). It is easy to see that $(\sum_ak_a^3)$ is a contact term because each component of $(\sum_a k_a^3)$ is independent  and therefore analytic in at least one of the momenta.

We now give an argument for why $N$ is likely to be suppressed 
 in the slow roll limit, so that the contribution to $\langle OOO\rangle '$ which arises from $S_h$ is sub-dominant compared to a solution of Ward identities with the $\langle OOOO\rangle $ source turned on, eq.(\ref{wardidsct}), eq.(\ref{wardidscal}).

To understand this point let us return to the canonical slow roll model. In this model, to leading order, no term of the form eq.(\ref{shfinal})
is present. One quick way to see this is to notice that in eq.(\ref{cusl}) there is no term of the form $(\sum_a k_a^3) \ln(\sum_b k_b)$. At subleading
order such a term does arise in this model, but it is suppressed with a coefficient of order $\epsilon^{3/2}$, as opposed to the leading terms in 
eq.(\ref{cusl}), which are $O(\sqrt{\epsilon})$. Having understood this better in the canonical model below, we will then argue that it should be true  more
generally as well, leading to the suppression of the $S_h$ contribution mentioned above.

In the canonical model, a term giving rise to a contribution of the form  eq.(\ref{shfinal}) would arise  from a contribution to  the Lagrangian of the form 
\be
\label{dels}
\int d^3x \,a^3 \, (V''' \delta \phi^{\,3}).
\ee
Comparing with eq.(3.8) in \cite{Maldacena:2002vr}, we see that such a contribution  is in fact present (in the second line). However, it is not included in the final result for the three point function  because it is suppressed. To keep the discussion simple  we assume that eq.(\ref{slcomp}) is valid,
and therefore that in the slow roll approximation every additional time derivative is suppressed with one factor of $\epsilon$, as was discussed in section \ref{basics}.
It is then straightforward to see that, barring accidental cancellations, this  requires every additional derivative of the potential to be suppressed by a factor of $\sqrt{\epsilon}$.

For example, from eq.(\ref{epsdelslr})   we see that 
\be
\label{condca}
{V' \over V} \sim  \sqrt{\epsilon}, ~ {V''\over V} \sim \epsilon,
\ee
so that 
\be
\label{rataa}
{V''\over V'}\sim \sqrt{\epsilon}.
\ee
Similarly, since eq.(\ref{sceq}) is valid, we have on taking two time derivatives
\be
\label{oderv}
\partial_t^3 \bar{\phi} \sim {V'''\over H} \, \dot{\bar{\phi}}^{\,2}.
\ee
Now
\be
\label{newrelaa}
\partial_t^3 \bar{\phi} \sim \epsilon^2 H^2 \dot{\bar{\phi}},
\ee
since the LHS has two additional time derivatives.
This gives, on using eq.(\ref{relasl}), 
\be
\label{valtrip}
{V''' \over H^2} \sim \epsilon^{3/2}.
\ee
So we see that $V'''$ (in units of $H^2$)  is smaller than the terms of order ${\dot{\bar{\phi}}\over H} \sim \sqrt{\epsilon}$, retained in eq.(\ref{cusl}).

In section \ref{moregenaction} towards the end, we argued that quite generically eq.(\ref{approxh}) and eq.(\ref{sceq}) are expected to be valid for a  general action of the form eq.(\ref{genlag})  in the slow roll approximation. 
It then follows, as was mentioned there, that every additional time derivative will be suppressed by one additional power of $\epsilon$, so that the argument above will go through, leading to eq.(\ref{valtrip}).

Let us end with some comments.  
First, if somehow due to say accidental cancellations,  the normalization constant $N$ is bigger than $O({\dot{\bar{\phi}}\over H})$, the three point function 
would be bigger in magnitude, making it more experimentally accessible. However, in this case if approximate conformal invariance is preserved, the functional form for $\langle OOO \rangle'$ must be as given by $S_h$,  eq.(\ref{shfinal}), and is completely fixed,  so this possibility can also be tested observationally.
Second,   by using the generalized Fourier transform discussed in appendix \ref{appSh}, we can write down a formal solution for the three point function in terms of the four point function. For completeness, we present this result  in  appendix \ref{prescription}. 
Finally, conformal perturbation theory is a standard way to study the  consequences of small departures from conformal invariance. In this, one perturbs  a conformally invariant theory  by turning on a coupling constant that breaks conformal invariance, and then calculates correlators perturbatively in this coupling constant.  Our approach above is different, and attempts to solve the Ward identities of scale and special conformal invariance after incorporating the effects of the breaking of these symmetries. This approach, which is akin to trying to solve the Callan-Symanzik equation for a small value of the beta function, can be more powerful in principle, although an explicit solution of the resulting Ward identities has not proved so easy in practice, as we see from appendix \ref{prescription}. 


\section{Conclusions}
\label{conclusions}

In this paper we have studied the constraints imposed by approximate conformal invariance on the scalar three point function. 
This correlation function is of the greatest interest experimentally, as a test of non-Gaussianity, and it is therefore important to understand how well it can be constrained in a model independent manner from symmetry considerations alone. 
In particular, we derived the Ward identities of scale and special conformal invariance and showed that these relate the three point function to the four point function in a particular limit, once the breaking of conformal invariance due to the non-zero values of slow roll parameters is taken into account. 

We then investigated these Ward identities and found that they considerably constrain the three point function. We argued that  as long as the dynamics is  approximately conformally invariant, and the slow roll approximation is valid, the magnitude of the three point function should be suppressed, being of the same order as that found in the canonical slow roll model of inflation, eq.(\ref{actslroll}).  Roughly, although the detailed functional form is different, this corresponds to $f_{NL}\sim O( ({\dot{\bar{\phi}}\over H})^2)$.  If an experimental discovery of non-Gaussianity is made in the near future it would almost certainly require a much bigger value for the three point correlator. Our analysis therefore says that such a discovery would not only rule out the canonical slow roll model of inflation, but in fact any model where conformal invariance is approximately valid, and the 
slow roll 
approximation is valid. 

We also found that the Ward identities determine the three point function in terms of the four point function nearly completely.  An    additional function, $S_h$, is allowed, but its functional form is completely fixed, and corresponds to the three point function of a dimension $3$  scalar primary operator in a CFT, only leaving the overall normalization and a coefficient of a contact term undetermined. We argued that generically  the overall  normalization should be suppressed  in the slow roll approximation. If somehow this generic argument fails and the normalization is bigger leading to $S_h$  dominating   the three point function, the functional form of the three point function would be  completely fixed, allowing for an experimental test of this possibility as well. 

 Unlike the three point function, the four point function does not vanish in the leading slow roll approximation, and is conformally invariant. By relating the three point function to the four point function we therefore relate the three point function also to a conformally invariant correlator. Unfortunately, as is well known, the functional form of the four point function is not constrained very significantly by conformal invariance alone; as a result  of the Ward identities this is also then true for the three point function.  In the canonical slow roll model the four point function arises due to single graviton exchange. If the three point function is observed and found to deviate from its functional form in the canonical slow roll 
model,  the four point function must also be different, suggesting  perhaps that 
higher spin fields are involved during inflation. This line of thought is well worth exploring further. 

More generally, it would be worth extending the analysis in this paper to include the breaking of conformal invariance to higher order in the slow roll expansion. 
The three point function, to leading non-vanishing order,  only requires corrections of order ${\dot{\bar{\phi}}\over H}$ to be included, and these can be obtained without changing the background geometry, since corrections to the metric are of order  the slow roll parameters, $\epsilon$ and $\delta$, eq.(\ref{defslowroll}),  and we have  argued that these should be much smaller. 
But going beyond  this order would require corrections   in the de Sitter geometry also to be incorporated. This is an interesting question to pursue, both from the point of view of cosmology and also holography in approximately AdS spaces. Once the asymptotic behavior of the fields has been determined, the Ward identities should follow from the invariance of the wave function under time and spatial reparametrizations. 


\section{Acknowledgements}
We  have benefited greatly from  discussions with  Suvrat Raju. SPT acknowledges support from the CERN  Theory Division for a sabbatical visit from July-December 2014 where some of this work was done. 
He also acknowledges support from the J. C. Bose fellowship of the Government of India and  from the Department of Atomic Energy, Government of India.
NK and SPT would like to acknowledge the hospitality of IISER Pune, and thank the organizers of the mini-workshop on string theory held there, for interesting discussions when this work was going on.
NK and AS would like to thank the organizers of the Advanced String School 2014, held at Puri, where part of this work was done.
Most of all, we thank the people of India for their generous support for research in string theory.

\appendix
\section{More on $\langle OOO\rangle$ and $\langle OOOO\rangle$ in the Canonical Model of Slow Roll Inflation}
\label{appcoeffn}
In this appendix, we discuss in some more detail the coefficient functions $\langle OOO\rangle$ and $\langle OOOO\rangle$ in the canonical model of slow roll inflation. We divide this appendix into two subsections, one for each of them.
\subsection{The Three Point Coefficient Function $\langle OOO\rangle$}
\label{appsubooo}
The three point scalar correlator $\langle \zeta(\vecs[k,1]) \zeta(\vecs[k,2]) 
\zeta(\vecs[k,3]) \rangle$ in the canonical slow roll model, eq.(\ref{actslroll}), was computed in \cite{Maldacena:2002vr},
\be
\langle \zeta(\vecs[k,1]) \zeta(\vecs[k,2]) \zeta(\vecs[k,3]) \rangle = (2 
\pi)^3 \delta^{3} \big(\vecs[k,1] + \vecs[k,2] + \vecs[k,3] \big) \, 
\frac{H^4}{\dot{\bar{\phi}}^4} 
\, \frac{H^4}{M_{Pl}^4} \, \frac{1}{\prod_{a} (2k_a^{\,3})} \, A,
\label{zzzmalda}
\ee
with
\be
A = \left( \frac{2 \ddot{\bar{\phi}}}{H 
\dot{\bar{\phi}}} + \frac{\dot{\bar{\phi}}^{\:2}}{2 H^2} 
\right) \big( \sum_{a} k_{a}^{\,3} \big)
+ \frac{\dot{\bar{\phi}}^{\:2}}{H^2} \left[ \half \sum_{a \neq b} 
k_{a} k_{b}^{2} + \frac{4}{k_{t}} \sum_{a>b} k_{a}^{2} k_{b}^{2} \right].
\label{astarmalda}
\ee
Here, $k_{a} = |\vecs[k,a]|$ and $k_{t} = k_1 + k_2 + k_3$.
Using the definitions of the slow-roll parameters, $\epsilon$ and $\delta$, 
eq.\eqref{defslowroll}, and the eq.\eqref{relasl}, in eq.\eqref{zzzmalda} and eq.\eqref{astarmalda} above, we can obtain the expression 
for $\langle \zeta(\vecs[k,1]) \zeta(\vecs[k,2]) \zeta(\vecs[k,3]) \rangle$ in 
terms of $\epsilon,~\delta$ as
\be
\langle \zeta(\vecs[k,1]) \zeta(\vecs[k,2]) \zeta(\vecs[k,3]) \rangle = (2 
\pi)^3 \delta^{3} \big( \vecs[k,1] + \vecs[k,2] + \vecs[k,3] \big) \, \frac{1}{4 \epsilon^2} 
\, \frac{H^4}{M_{Pl}^4} \, \frac{1}{\prod_{a} (2k_a^{\,3})} \, A,
\label{zzzsroll}
\ee
with
\be
A = \left( \epsilon + 2\delta \right) \big( \sum_{a} k_{a}^{\,3} 
\big)
+  2 \, \epsilon \left[ \half \sum_{a \neq b} k_{a} k_{b}^{2} + 
\frac{4}{k_{t}} \sum_{a>b} k_{a}^{2} k_{b}^{2} \right].
\label{astarsroll}
\ee

We can also express the relation between $\zeta$ and $\delta\phi$, as given in eq.\eqref{relzpb}, in terms of the parameters $\epsilon$ and $\delta$ as
\be
\zeta = - \, \frac{1}{\sqrt{2\epsilon}} \, \delta\phi - 
\left(\frac{\epsilon + \delta}{4\epsilon} \right) 
\delta\phi^2.
\ee
Then from eq.\eqref{zzzsroll} and eq.\eqref{astarsroll}, we get
\begin{equation}
\begin{split}
\langle \delta\phi(\vecs[k,1]) \delta\phi(\vecs[k,2]) \delta\phi(\vecs[k,3]) 
\rangle &= - \, (2 \pi)^3 \delta^{3} \big(\vecs[k,1] + \vecs[k,2] + \vecs[k,3] \big) \,\frac{H^4}{M_{Pl}^4} \, 
\frac{1}{\prod_{a} (2k_a^{\,3})} \, \times \\ &\bigg[ \bigg( \frac{3\epsilon + 4\delta}{\sqrt{2\epsilon}} \bigg)
\sum_{a} k_{a}^{\,3} + \sqrt{2\epsilon} \, \bigg( \half \sum_{a \neq b} k_a 
k_b^2 + \frac{4}{k_t} \sum_{a>b} k_a^2 k_b^2 \bigg) \bigg].
\label{dphicube}
\end{split}
\end{equation}

Now, to obtain a relationship between $\langle \delta\phi(\vecs[k,1]) \delta\phi(\vecs[k,2]) \delta\phi(\vecs[k,3]) \rangle$ and $\langle 
O(\vecs[k,1]) O(\vecs[k,2]) O(\vecs[k,3]) \rangle$, we use the momentum space expression for the wave function eq.\eqref{wf1}, given by
\begin{align}
\begin{split}
\psi[\delta\phi] = \text{exp} \bigg[ \frac{M_{Pl}^2}{H^2} \bigg( &-\frac{1}{2!} \int \frac{d^{3}\vecs[k,1]}{(2\pi)^3} \frac{d^{3}\vecs[k,2]}{(2\pi)^3} \, \delta\phi(\vecs[k,1]) \delta\phi(\vecs[k,2]) \, \langle O(-\vecs[k,1]) O(-\vecs[k,2]) \rangle \\
 &+ \frac{1}{3!} \int \frac{d^{3}\vecs[k,1]}{(2\pi)^3} \frac{d^{3}\vecs[k,2]}{(2\pi)^3} \frac{d^{3}\vecs[k,3]}{(2\pi)^3} \, \delta\phi(\vecs[k,1]) \delta\phi(\vecs[k,2]) \delta\phi(\vecs[k,3]) \, \times \\ & \hspace{40mm} \langle O(-\vecs[k,1]) O(-\vecs[k,2]) O(-\vecs[k,3]) \rangle \bigg) \bigg],
\label{wavefmomsp}
\end{split}
\end{align}
where we have kept only the relevant terms. This gives
\be
\langle \delta\phi(\vecs[k,1]) \delta\phi(\vecs[k,2]) \delta\phi(\vecs[k,3]) 
\rangle = \quarter \, \frac{H^4}{M_{Pl}^4} \, \frac{\langle O(\vecs[k,1]) 
O(\vecs[k,2]) O(\vecs[k,3]) \rangle}{\prod_{a=1}^{3} \langle O(\vecs[k,a]) 
O(-\vecs[k,a]) \rangle{'}}\,.
\label{ooophi3}
\ee
Using the expression for $\langle O(\vecs[k,a]) O(-\vecs[k,a]) \rangle{'}$, eq.\eqref{defoo}, in eq.\eqref{ooophi3}, and using eq.\eqref{dphicube} we obtain the relation 
\be
\langle O(\vecs[k,1]) O(\vecs[k,2]) O(\vecs[k,3]) \rangle' = - \, {3\epsilon + 4\delta \over {2 \sqrt{2\epsilon}}} \sum_{a} k_{a}^{\,3} - \half \sqrt{2\epsilon} \, \bigg( \half \sum_{a \neq b} k_a k_b^2 + \frac{4}{k_t} \sum_{a>b} k_a^2 k_b^2 \bigg),
\ee
which is same as the expression in eq.\eqref{cusl}.
\subsection{The Four Point Coefficient Function $\langle OOOO\rangle$}
\label{appsuboooo}
The scalar four point coefficient function $\langle OOOO \rangle$ in the canonical slow roll model was calculated  in \cite{Seery:2008ax} and \cite{Ghosh:2014kba}. It is given, see eq.(4.32) of \cite{Ghosh:2014kba}, as 
\be
\label{valopos}
\langle O({\vect{x}}_1) O({\vect{x}}_2) O({\vect{x}}_3) O({\vect{x}}_4)\rangle =\int \prod_{a=1}^4 {d^3 k_a \over (2\pi)^3}
 e^{i \vecs[k,a] \cdot \vecs[x,a]} \, \langle O(\vk) O(\vkk) O(\vkkk) O(\vkfour)\rangle \, ,
\ee
where
\be
\label{fourptwoapp}
\begin{split}
\langle O(\vk)& O(\vkk) O(\vkkk) O(\vkfour)\rangle =-4 \, (2 \pi)^3 \delta^3(\sum_{a=1}^4
\vecs[k,a]) \bigg[ {1\over2}\bigg\{\widehat{W}^S(\vk, \vkk, \vkkk, \vkfour)
\\&+\widehat{W}^S(\vk, \vkkk, \vkk, \vkfour)+\widehat{W}^S(\vk, \vkfour, \vkkk, \vkk)\bigg\} +\widehat{R}^S(\vk, \vkk, \vkkk, \vkfour) \\
&+\widehat{R}^S(\vk, \vkkk, \vkk, \vkfour) +\widehat{R}^S(\vk, \vkfour, \vkkk, \vkk)\bigg],
\end{split}
\ee
with ${\widehat W}^S$ being the contribution from the transverse component of the graviton exchanged, given by
{\small
\begin{align}
\label{hatW}
\begin{split}
 &\widehat{W}^S(\vk,\vkk,\vkkk,\vkfour) = - 2 \bigg[ \bigg\{\vk.\vkkk+\frac{\{(\vkk+\vk).\vk\} \{(\vkfour+\vkkk).\vkkk\}}{|\vk+\vkk|^2}\bigg\} \\ & \bigg\{\vkk.\vkfour+  \frac{\{(\vk+\vkk).\vkk\} \{(\vkkk+\vkfour).\vkfour\}}{|\vk+\vkk|^2}\bigg\} 
+  \bigg\{\vk.\vkfour+\frac{\{(\vkk+\vk).\vk\} \{(\vkfour+\vkkk).\vkfour\}}{|\vk+\vkk|^2}\bigg\}\\ & 
\bigg\{\vkk.\vkkk+\frac{\{(\vkk+\vk).\vkk\} \{(\vkfour+\vkkk).\vkkk\}} {|\vk+\vkk|^2}\bigg\}-
\bigg\{\vk.\vkk-\frac{\{(\vkk+\vk).\vk\} \{(\vk+\vkk).\vkk\}} {|\vk+\vkk|^2}\bigg\}\\ & 
\bigg\{\vkkk.\vkfour-\frac{\{(\vkkk+\vkfour).\vkfour\} \{(\vkfour+\vkkk).\vkkk\}}{|\vk+\vkk|^2}\bigg\} \bigg] \times\\ & 
\Bigg[ \bigg\{\frac{ k_{1} k_{2} (k_{1}+k_{2})^2 \left((k_{1}+k_{2})^2-k_{3}^2-k_{4}^2-4 k_{3} k_{4}\right)}{ (k_{1}+k_{2}-k_{3}-k_{4})^2 (k_{1}+k_{2}+k_{3}+k_{4})^2
   (k_{1}+k_{2}-|\vk+\vkk|) (k_{1}+k_{2}+|\vk+\vkk|)} \\ &\Big(-\frac{k_{1}+k_{2}}{2 k_{1}
   k_{2}}-\frac{k_{1}+k_{2}}{-(k_{1}+k_{2})^2+k_3^2+k_{4}^2+4 k_{3}
   k_{4}}+\frac{k_{1}+k_{2}}{|\vk+\vkk|^2-(k_{1}+k_2)^2}\\ &+\frac{1}{-k_{1}-k_{2}+k_{3}+k_{4}} -\frac{1}{k_{1}+k_{2}+k_{3}+k_{4}}+\frac{3}{2 (k_{1}+k_{2})}\Big)  + (1,2 \leftrightarrow 3,4) \bigg\}\\
&- \frac{|\vk+\vkk|^3 \left(-k_{1}^2-4 k_{2}
   k_{1}-k_{2}^2+|\vk+\vkk|^2\right) \left(-k_{3}^2-4
   k_{4} k_{3}-k_{4}^2+|\vk+\vkk|^2\right)}{2
   \left(-k_{1}^2-2 k_{2}
   k_{1}-k_{2}^2+|\vk+\vkk|^2\right)^2 \left(-k_{3}^2-2
   k_{4} k_{3}-k_{4}^2+|\vk+\vkk|^2\right)^2}\Bigg].
\end{split}
\end{align} }
The longitudinal contribution from the graviton is denoted by $\widehat{R}^S$, and is given by
{\small
\begin{equation} \label{remexp}
\widehat{R}^S(\vk,\vkk,\vkkk,\vkfour) = {A_1(\vk,\vkk,\vkkk,\vkfour) \over (k_1+k_2+k_3+k_4)}  + {A_2(\vk,\vkk,\vkkk,\vkfour) \over (k_1+k_2+k_3+k_4)^2} + {A_3(\vk,\vkk,\vkkk,\vkfour) \over (k_1+k_2+k_3+k_4)^3}
\end{equation}}
with
{\small
\begin{align} \nonumber
  A_1(\vk,\vkk,\vkkk,\vkfour) = & \bigg[ \frac{\vkkk \cdot \vkfour \left(\vk \cdot \vkk \left(k_1^2+k_2^2\right)+2 k_1^2 k_2^2\right)}{8 |\vk+\vkk|^2} + \{1,2 \Leftrightarrow 3,4\}\bigg] \\ \nonumber &-\frac{k_1^2 \vkk \cdot \vkkk k_4^2+k_1^2 \vkk \cdot \vkfour k_3^2+\vk \cdot \vkkk k_2^2 k_4^2+\vk \cdot \vkfour k_2^2 k_3^2}{2 |\vk+\vkk|^2} \\ &-\frac{\left(\vk \cdot \vkk \left(k_1^2+k_2^2\right)+2 k_1^2 k_2^2\right) \left(\vkkk \cdot \vkfour \left(k_3^2+k_4^2\right)+2 k_3^2 k_4^2\right)}{8 |\vk+\vkk|^4}, \label{A1inR}  
  \end{align}
  \begin{align} \nonumber
  A_2(&\vk,\vkk,\vkkk,\vkfour) =  -\frac{1}{8 |\vk+\vkk|^4}\bigg[k_3 k_4 (k_3+k_4) \left(\vk \cdot \vkk \left(k_1^2+k_2^2\right)+2 k_1^2 k_2^2\right)\\  \nonumber & (k_3 k_4+\vkkk \cdot \vkfour) \nonumber  +k_1 k_2 (k_1+k_2) (k_1 k_2+\vk \cdot \vkk) \left(\vkkk \cdot \vkfour \left(k_3^2+k_4^2\right)+2 k_3^2 k_4^2\right)\bigg] \\&  
  -\frac{1}{2 |\vk+\vkk|^2}\bigg[k_1^2 \vkk \cdot \vkkk k_4^2 (k_2+k_3)+k_1^2 \vkk \cdot \vkfour k_3^2 (k_2+k_4) \nonumber \\ &+\vk \cdot \vkkk k_2^2 k_4^2 (k_1+k_3)+\vk \cdot \vkfour k_2^2 k_3^2 (k_1+k_4)\bigg] \nonumber  \\ & + \bigg[\frac{\vk \cdot \vkk}{8 |\vk+\vkk|^2} \big((k_1+k_2) \left(\vkkk \cdot \vkfour \left(k_3^2+k_4^2\right)+2 k_3^2 k_4^2\right) \nonumber \\
  & + k_3 k_4 (k_3+k_4) (k_3 k_4+\vkkk \cdot \vkfour)\big)  + \{1,2 \Leftrightarrow 3,4\} \bigg],\label{A2inR}
  \end{align}
  \begin{align} \nonumber
  A_3(\vk&,\vkk,\vkkk,\vkfour) =  -\frac{k_1 k_2 k_3 k_4 (k_1+k_2) (k_3+k_4) (k_1 k_2+\vk \cdot \vkk) (k_3 k_4+\vkkk \cdot \vkfour)}{4 |\vk+\vkk|^4}\\ \nonumber & 
 -\frac{k_1 k_2 k_3 k_4 (k_1 \vkk \cdot \vkkk k_4+k_1 \vkk \cdot \vkfour k_3+\vk \cdot \vkkk k_2 k_4+\vk \cdot \vkfour k_2 k_3)}{|\vk+\vkk|^2}\\ & + \frac{1}{4 |\vk+\vkk|^2} \bigg[k_1 k_2 (k_1 k_2+\vk \cdot \vkk) \left(\vkkk \cdot \vkfour \left(k_3^2+k_4^2\right)+2 k_3^2 k_4^2\right) \nonumber \\ &+\vk \cdot \vkk k_3 k_4 (k_1+k_2) (k_3+k_4) (k_3 k_4+\vkkk \cdot \vkfour)+ \{1,2 \Leftrightarrow 3,4\}\bigg] \nonumber \\ & +\frac{3 k_1 k_2 k_3 k_4 (k_1 k_2+\vk \cdot \vkk) (k_3 k_4+\vkkk \cdot \vkfour)}{4 |\vk+\vkk|^2}. \label{A3inR}
\end{align}}
From eq.(\ref{fpzeta}), we can see that $\langle \zeta \zeta \zeta \zeta \rangle$ is made up of two parts. Among them, $\langle \zeta \zeta \zeta \zeta\rangle_{CF}$ gets contribution from the four point coefficient function $\langle OOOO \rangle$. Similar to eq.(\ref{ooophi3}), one can derive a relation between $\langle OOOO \rangle$ and $\langle \delta\phi \, \delta\phi \, \delta\phi \, \delta\phi \rangle_{CF}$ using the momentum space wave function. The relation is given by
\be
\langle \delta\phi(\vecs[k,1]) \delta\phi(\vecs[k,2]) \delta\phi(\vecs[k,3]) \delta\phi(\vecs[k,4])
\rangle_{CF} = {1 \over 8} \, \frac{H^6}{M_{Pl}^6} \, \frac{\langle O(\vecs[k,1]) 
O(\vecs[k,2]) O(\vecs[k,3])O(\vecs[k,4]) \rangle}{\prod_{a=1}^{4} \langle O(\vecs[k,a]) 
O(-\vecs[k,a]) \rangle{'}}.
\label{oooophi4}
\ee
Inverting eq.(\ref{relzpb}), we obtain $\delta\phi$ in terms of $\zeta$. Working upto linear order in $\delta\phi$, we get
\be
\label{dpzetalin}
\delta \phi = -\,{\dot{\bar{\phi}} \over H} \, \zeta\,.
\ee
Using eq.(\ref{dpzetalin}) in eq.(\ref{oooophi4}), we obtain 
\begin{equation}
\label{fpzetacf}
 \langle \zeta(\vecs[k,1]) \zeta(\vecs[k,2]) \zeta(\vecs[k,3]) \zeta(\vecs[k,4]) \rangle_{CF} = {1 \over 8} \, {H^6 \over M_{Pl}^6} \, {H^4 \over {{\dot{\bar{\phi}}}}^4}\, {\langle O(\vecs[k,1]) O(\vecs[k,2]) O(\vecs[k,3]) O(\vecs[k,4]) \rangle \over \prod_{a=1}^4 \langle O(\vecs[k,a])  O(-\vecs[k,a])\rangle'}
\end{equation}

Similarly, the other contribution in $\langle \zeta \zeta \zeta \zeta \rangle$, i.e. $\langle \zeta \zeta \zeta \zeta \rangle_{ET}$, comes from integrating out a 
boundary graviton. The corresponding $\langle \delta\phi \, \delta\phi \, \delta\phi \, \delta\phi \rangle_{ET}$ was computed in eq.(5.6) of \cite{Ghosh:2014kba},

\begin{align}
\label{valGE}
\begin{split}
\langle \delta \phi({\bf k}_1) \delta\phi({\bf k}_2) &\delta \phi({\bf k}_3) \delta\phi({\bf k}_4) \rangle_{ET}= 4 (2 \pi)^3 \delta^3\Big(\sum_{a=1}^4 \vecs[k,a] \Big) {H^6 \over M_{Pl}^6} \, {1 \over \prod_{a=1}^4 (2 k_a^{\,3})} \\ & \bigg[\widehat{G}^S({\bf k}_1, {\bf k}_2, {\bf k}_3, {\bf k}_4) +\widehat{G}^S({\bf k}_1, {\bf k}_3, {\bf k}_2, {\bf k}_4) +\widehat{G}^S({\bf k}_1, {\bf k}_4, {\bf k}_3, {\bf k}_2)\bigg],
\end{split}
\end{align}
with $\widehat{G}^S$ being given 
by (eq.(5.7) of \cite{Ghosh:2014kba})
\be
\label{defGa} {\small 
\begin{split}
&\widehat{G}^S({\bf k}_1, {\bf k}_2, {\bf k}_3, {\bf k}_4) = {S(\widetilde{{\bf k}},{\bf k}_1, {\bf k}_2)S(\widetilde{{\bf k}},{\bf k}_3, {\bf k}_4)  \over |{\bf k}_1+{\bf k}_2|^3}\bigg[ \bigg\{{\bf k}_1.{\bf k}_3+\frac{\{({\bf k}_2+{\bf k}_1).{\bf k}_1\} \{({\bf k}_4+{\bf k}_3).{\bf k}_3\}}{|{\bf k}_1+{\bf k}_2|^2}\bigg\}\\ & \bigg\{{\bf k}_2.{\bf k}_4+  \frac{\{({\bf k}_1+{\bf k}_2).{\bf k}_2\} \{({\bf k}_3+{\bf k}_4).{\bf k}_4\}}{|{\bf k}_1+{\bf k}_2|^2}\bigg\} 
+  \bigg\{{\bf k}_1.{\bf k}_4+\frac{\{({\bf k}_2+{\bf k}_1).{\bf k}_1\} \{({\bf k}_4+{\bf k}_3).{\bf k}_4\}}{|{\bf k}_1+{\bf k}_2|^2}\bigg\}\\ & 
\bigg\{{\bf k}_2.{\bf k}_3+\frac{\{({\bf k}_2+{\bf k}_1).{\bf k}_2\} \{({\bf k}_4+{\bf k}_3).{\bf k}_3\}} {|{\bf k}_1+{\bf k}_2|^2}\bigg\}-
\bigg\{{\bf k}_1.{\bf k}_2-\frac{\{({\bf k}_2+{\bf k}_1).{\bf k}_1\} \{({\bf k}_1+{\bf k}_2).{\bf k}_2\}} {|{\bf k}_1+{\bf k}_2|^2}\bigg\}\\ & 
\bigg\{{\bf k}_3.{\bf k}_4-\frac{\{({\bf k}_3+{\bf k}_4).{\bf k}_4\} \{({\bf k}_4+{\bf k}_3).{\bf k}_3\}}{|{\bf k}_1+{\bf k}_2|^2}\bigg\} \bigg],
\end{split}}
\ee with
\be
\label{singhat}
S( \widetilde{{\bf k}},{\bf k}_1, {\bf k}_2) = (k_1+k_2+k_3) - {\sum_{i>j}k_ik_j \over (k_1+k_2+k_3)} - {k_1k_2k_3\over (k_1+k_2+k_3)^2} \Bigg|_{{\bf k}_3 \, = \, \widetilde{{\bf k}} \, = \, -\,({\bf k}_1+{\bf k}_2)}.
\ee

 In eq.(\ref{valGE}), one can use eq.(\ref{dpzetalin}) to obtain
\begin{align}
\label{fpzetaet}
\begin{split}
 \langle \zeta(\vecs[k,1]) \zeta(\vecs[k,2]) &\zeta(\vecs[k,3]) \zeta(\vecs[k,4]) \rangle_{ET} = 4 \, (2\pi)^3\delta^3\Big(\sum_{a=1}^4 \vecs[k,a]\Big) \, {H^4 \over {{\dot{\bar{\phi}}}}^4} \, {H^6 \over M_{Pl}^6} \, {1 \over \prod_{a=1}^4 (2 k_a^{\,3})} \, \times \\
& \bigg[\widehat{G}^S(\vk, \vkk, \vkkk, \vkfour) +\widehat{G}^S(\vk, \vkkk, \vkk, \vkfour) +\widehat{G}^S(\vk, \vkfour, \vkkk, \vkk)\bigg].
 \end{split}
\end{align}

Thus, $\langle \zeta(\vecs[k,1]) \zeta(\vecs[k,2]) \zeta(\vecs[k,3]) \zeta(\vecs[k,4]) \rangle_{CF}$, in eq.(\ref{fpzetacf}), and $\langle \zeta(\vecs[k,1]) \zeta(\vecs[k,2]) \zeta(\vecs[k,3]) \zeta(\vecs[k,4]) \rangle_{ET}$, in eq.(\ref{fpzetaet}), give the two contributions mentioned on the RHS. of eq.(\ref{fpzeta}).

\section{Solving the Homogeneous Equation for $\langle OOO\rangle$}
\label{appSh}
In this appendix, we calculate the homogeneous contribution to the three point function $\langle OOO \rangle'$, denoted by $S_h(\vecs[k,1],\vecs[k,2],\vecs[k,3])$, eq.\eqref{defSh}. For this, we need to solve the equations eq.\eqref{homeqscale} and eq.\eqref{homeqsct}. We start by rewriting eq.\eqref{homeqsct} in a slightly different manner which is more suited for the purpose of our calculation. Note that the function $S_h(\vecs[k,1],\vecs[k,2],\vecs[k,3])$ is a function only of the magnitudes $k_1, k_2$ and $k_3$.
Thus it will be beneficial for us if we express the derivative operators in eq.\eqref{delom} in terms of the magnitudes $k_1, k_2$ and $k_3$, rather then in terms of the components of $\vecs[k,1], \vecs[k,2] \: \text{and} \: \vecs[k,3]$.
Using
\begin{align}
\frac{\partial}{\partial {k_i}} = \frac{k_i}{k} \, \frac{\del}{\del k} \,,\label{dcompmag1}
\end{align}
where $k$ is the magnitude and $k_i$ is the $i^{th}$ component of a generic vector $\vec[k]$, we can re-express the derivative operator ${\cal L}^{\vec[b]}_{\vec[k]}$ as
\begin{align}
\label{deltao4}
{\cal L}^{\vec[b]}_{\vec[k]} &= (\vec[b]\cdot\vec[k]) \, \Theta(k)
\end{align}
with
\begin{align}
\Theta(k) = -\,\frac{2}{k} \, \frac{\del}{\del k} + \frac{\del^{\,2}}{\del k^{2}}\,.
\label{deftheta}
\end{align}
Eq.\eqref{homeqsct} can then be written as
\be
\big[ (\vec[b]\cdot\vecs[k,1]) \Theta(k_1) + (\vec[b]\cdot\vecs[k,2]) \Theta(k_2) + (\vec[b]\cdot\vecs[k,3]) \Theta(k_3) \big] \, S_h(k_1, k_2, k_3) = 0.
\label{wardhomo6}
\ee
With the choice for the parameter of the special conformal transformation, $\vec[b]$, to be perpendicular to $\vecs[k,3]$, \emph{i.e.} $\vec[b] \perp \vecs[k,3]$, eq.\eqref{wardhomo6}  becomes
\be
(\Theta(k_1) - \Theta(k_2)) \, S_h(k_1, k_2, k_3) = 0.
\label{th1th2s}
\ee
Similarly, we can make another independent choice for the parameter $\vec[b]$, $\vec[b] \perp \vecs[k,2]$, and obtain
\be
(\Theta(k_1) - \Theta(k_3)) \, S_h(k_1, k_2, k_3) = 0.
\label{th1th3s}
\ee
The  other possible independent choice, $\vec[b] \perp \vecs[k,1]$, gives an  equation that is a linear combination of eq.\eqref{th1th2s} and eq.\eqref{th1th3s}.

We will now analyze solutions to these equations. Our analysis is related to that carried out in \cite{Mata:2012bx}.
Let us consider a complete set of functions $f_{z}(k)$ defined in the range $z \in (-\infty,\infty)$, given by
\be
f_{z}(k) = (1+ikz)\,e^{-ikz}.
\label{deffzk}
\ee
Any general function, say $H(k)$, can be expanded in terms of $f_z(k)$ in a souped-up Fourier transform as 
\be
\label{defsf}
H(k)= \int_{-\infty}^{\infty} dz \, f_z(k) \, \tilde{H}(z).
\ee
The functions $f_{z}(k)$ are actually eigenfunctions of the operators $\Theta(k)$, satisfying 
\be
\Theta(k)f_{z}(k) = - z^2 f_{z}(k).
\label{thetaeigen}
\ee
It is also important to note that the inverse of the transformation in eq.\eqref{defsf} is given by,
\be
\label{invsf}
\tilde{H}(z) =- \int_{-\infty}^{\infty} {dk \over 2\pi} \, \bigg(k \, e^{i k z} \int^k {H(q) \over q^2} dq \bigg).
\ee
Using eq.\eqref{defsf}, we can expand the function $S_h(k_1, k_2, k_3)$ as
\be
S_h(k_1, k_2, k_3) = \int_{-\infty}^{\infty} dz_1 \, dz_2 \, dz_3 \, f_{z_1}(k_1) f_{z_2}(k_2) f_{z_3}(k_3) \, \mathcal{M}(z_1,z_2,z_3).
\label{sfoursoup}
\ee
Substituting $S_h(k_1, k_2, k_3)$ from eq.\eqref{sfoursoup} into eq.\eqref{th1th2s} and eq.\eqref{th1th3s}, we obtain
\be
z_1^{\, 2} = z_2^{\, 2} = z_3^{\, 2},
\ee
which in turn allows us to write $S_h(k_1, k_2, k_3)$ as
\be
S_h(k_1, k_2, k_3) = \sum_{n_1,n_2,n_3 = \pm 1} \int_{0}^{\infty} dz \, \mathcal{F}_{n_1 n_2 n_3}(k_1, k_2, k_3, z) \, \mathcal{M}_{n_1 n_2 n_3}(z),
\label{sfm}
\ee
where $\mathcal{M}_{n_1 n_2 n_3}(z)$ are a set of 8 functions corresponding to the 8 possible choices of the set $\lbrace n_1, n_2, n_3\rbrace$, and $\mathcal{F}_{n_1 n_2 n_3}(k_1, k_2, k_3, z)$ is given by
\be
\mathcal{F}_{n_1 n_2 n_3}(k_1, k_2, k_3, z) = (1 + i n_1 k_1 z)\,(1 + i n_2 k_2 z)\,(1 + i n_3 k_3 z)\,e^{-i (n_1 k_1+n_2 k_2+n_3 k_3) z}.
\label{funcf}
\ee
Using eq.\eqref{dcompmag1}, we can also rewrite eq.\eqref{homeqscale} as
\be
\bigg( k_1 \frac{\partial}{\partial k_1} + k_2 \frac{\partial}{\partial k_2} + k_3 \frac{\partial}{\partial k_3} \bigg) S_h(k_1, k_2, k_3) = 3 \, S_h(k_1, k_2, k_3).
\label{dilacons1}
\ee
Using eq.\eqref{sfm} and eq.\eqref{funcf} in eq.\eqref{dilacons1} we get
\begin{align}
\begin{split}
\bigg[ \sum_{a=1}^3 k_a \, \frac{\partial}{\partial k_a} \bigg] S_h(k_1, k_2, k_3) = - \sum_{n_1,n_2,n_3 = \pm 1} \int_{0}^{\infty} dz \, \mathcal{F}_{n_1 n_2 n_3}&(k_1, k_2, k_3, z) \, \, \times \\ & \frac{\partial}{\partial z} \bigg[ z \, \mathcal{M}_{n_1 n_2 n_3}(z) \bigg].
\label{dilacons2}
\end{split}
\end{align}
Combining eq.\eqref{dilacons1} and eq.\eqref{dilacons2} we obtain
\be
\frac{\partial}{\partial z}\bigg[ z \, \mathcal{M}_{n_1 n_2 n_3}(z) \bigg] + 3 \, \mathcal{M}_{n_1 n_2 n_3}(z) = 0.
\ee
This has the general solution
\be
\mathcal{M}_{n_1 n_2 n_3}(z) = \frac{m_{n_1 n_2 n_3}}{z^4} \, ,
\label{mfuncz}
\ee
where $m_{n_1 n_2 n_3}$ is a $z$ independent constant. Thus, eq.\eqref{mfuncz} fixes the functional dependence of $\mathcal{M}$ on $z$.
Using eq.\eqref{mfuncz} in eq.\eqref{sfm} we see that
\be
S_h(k_1, k_2, k_3) = \sum_{n_1,n_2,n_3 = \pm 1} m_{n_1 n_2 n_3} \int_{0}^{\infty} \frac{dz}{z^4} \, \mathcal{F}_{n_1 n_2 n_3}(k_1, k_2, k_3, z).
\label{sintf}
\ee
To make the integration in eq.\eqref{sintf} well defined as $z \rightarrow \infty$, we add a small imaginary component to $k_a$. The integral is also divergent as $z \rightarrow 0$. We regularize it by putting a small cut-off  at $z = \lambda$. On carrying out the integral we get

\begin{align}
\begin{split}
S_h(k_1, k_2, k_3) = &\sum_{n_1,n_2,n_3 = \pm 1} m_{n_1 n_2 n_3} \Bigg\lbrace \frac{1}{3\lambda^3} + \frac{1}{2\lambda} \sum_{a=1}^3 (n_a k_a)^2 + \\
&+ i \Big(-\frac{4}{9} \sum_{a=1}^3 (n_a k_a)^3 -\frac{1}{3} \sum_{a \neq b} n_a k_a (n_b k_b)^2 + \frac{1}{3} \prod_{a=1}^3 (n_a k_a) \Big) \\
&- \frac{i}{3} \Big(\sum_{a=1}^3 (n_a k_a)^3 \Big) \Big( \int_{\lambda}^{\infty} \frac{dz}{z} e^{-i(n_1 k_1 + n_2 k_2 + n_3 k_3)z} \Big) \Bigg\rbrace.
\label{sfinal}
\end{split}
\end{align}
This gives us the solution to the homogeneous equations eq.\eqref{homeqscale} and eq.\eqref{homeqsct}. At this stage, it consists of a sum of eight distinct functions, corresponding to the eight distinct choices for the set $(n_1,n_2,n_3)$. We will now take various limits of the answer in eq.\eqref{sfinal} and find a unique solution. 

First of all, we remove the first two terms in the solution eq.\eqref{sfinal} which go like powers of $1/\lambda$, since their presence would violate conformal invariance. We next consider the last term involving the integral. We can explicitly evaluate this integral to get
\begin{align}
\int_\lambda^\infty \frac{dz}{z} \, e^{-\,i \, (\sum_{a} n_a k_a) \, z} = \Gamma[0,i (\sum_{a} n_a k_a) \lambda] = -\,\gamma -\frac{i\pi}{2} - ln\big[\lambda \big(\sum_{a} n_a k_a\big)\big] + O(\lambda).
\label{intexp}
\end{align}
Here, $\gamma$ is the Euler-Mascheroni constant and $ln$ denotes the natural logarithm. The $O(\lambda)$ terms appearing in eq.\eqref{intexp} vanish in the limit $\lambda \rightarrow 0$. Thus, our answer becomes
\begin{align}
\begin{split}
S_h(k_1, k_2, k_3) = \sum_{n_1,n_2,n_3 = \pm 1} m_{n_1 n_2 n_3} \Bigg\lbrace \frac{i}{3} \Big(\sum_{a=1}^3 (n_a k_a)^3\Big) \Big( \gamma + \frac{i\pi}{2} + ln\big[\lambda \big(\sum_{a} n_a k_a\big)\big] \Big) \\ + \, i \Big(-\frac{4}{9} \sum_{a=1}^3 (n_a k_a)^3 -\frac{1}{3} \sum_{a \neq b} n_a k_a (n_b k_b)^2 + \frac{1}{3} \prod_{a=1}^3 (n_a k_a) \Big) \Bigg\rbrace.
\label{ssimp}
\end{split}
\end{align}
We will now consider the behavior of eq.\eqref{ssimp} in the limit $k_1 \approx k_2 \gg k_3$. We know that the momentum space three point function is related to the position space expression by the standard Fourier transform. Thus
\begin{align}
\langle O(\vecs[k,1]) O(\vecs[k,2]) O(\vecs[k,3])\rangle &= \int d^3x_1 \, d^3x_2 \, d^3x_3 \, e^{-\,i (\sum_{a} \vecs[k,a]\cdot\vecs[x,a])} \langle O(\vecs[x,1]) O(\vecs[x,2]) O(\vecs[x,3])\rangle \nonumber \\ &= \int d^3x_1 \, d^3x_2 \, d^3x_3 \,e^{-\,i\lbrace(\vecs[k,1]+\vecs[k,2]+\vecs[k,3])\cdot\vecs[x,1] + \vecs[k,2]\cdot(\vecs[x,2]-\vecs[x,1])+\vecs[k,3]\cdot(\vecs[x,3]-\vecs[x,1]) \rbrace} \nonumber \\ &\hspace{52mm} \langle O(0) O(\vecs[x,2]-\vecs[x,1]) O(\vecs[x,3]-\vecs[x,1])\rangle \nonumber \\ &= \int d^3x_1 \, d^3x \, d^3y \, e^{-\,i\lbrace(\vecs[k,1]+\vecs[k,2]+\vecs[k,3])\cdot\vecs[x,1] + \vecs[k,2]\cdot\vec[x]+\vecs[k,3]\cdot\vec[y] \rbrace} \langle O(0) O(\vec[x]) O(\vec[y])\rangle \nonumber\\ &= (2\pi)^3 \, \delta^3\big(\sum_{a=1}^3 \vecs[k,a]\big) \int d^3x \, d^3y \, e^{-\,i(\vecs[k,2]\cdot\vec[x] + \vecs[k,3]\cdot\vec[y])} \langle O(0) O(\vec[x]) O(\vec[y])\rangle,
\label{tpft}
\end{align}
where we have used the notation $\vecs[x,2]-\vecs[x,1] = \vec[x]$ and $\vecs[x,3]-\vecs[x,1] = \vec[y]$. Now, as we are interested in the limit $k_2 \rightarrow \infty \Rightarrow x \rightarrow 0$ (where $x \equiv |\vec[x]|$), we can use the Operator Product Expansion (OPE)
\be
O(0) O(\vec[x]) = \frac{A}{x^3} \, O(\vec[x]) + \dots
\label{scalarope}
\ee
where $A$ is a constant. Substituting eq.\eqref{scalarope} into eq.\eqref{tpft} then gives us
\begin{align}
\begin{split}
\langle O(\vecs[k,1]) O(\vecs[k,2]) O(\vecs[k,3])\rangle &\approx (2\pi)^3 \, \delta^3\big(\sum_{a=1}^3 \vecs[k,a]\big) \int d^3x \, d^3y \, e^{-\,i(\vecs[k,2]\cdot\vec[x] + \vecs[k,3]\cdot\vec[y])} \, \frac{1}{x^3} \, \langle O(\vec[x]) O(\vec[y])\rangle \\ &= (2\pi)^3 \, \delta^3\big(\sum_{a=1}^3 \vecs[k,a]\big) \int d^3x \, d^3y \, e^{-\,i(\vecs[k,2]\cdot\vec[x] + \vecs[k,3]\cdot\vec[y])} \, \frac{1}{x^3} \, \frac{1}{|\vec[x] - \vec[y]|^6} \\ &\approx (2\pi)^3 \, \delta^3\big(\sum_{a=1}^3 \vecs[k,a]\big) \int d^3x \, d^3y \, e^{-\,i(\vecs[k,2]\cdot\vec[x] + \vecs[k,3]\cdot\vec[y])} \, \frac{1}{x^3} \, \frac{1}{y^6}\,,
\end{split}
\label{tpoplim}
\end{align}
where we have used the fact that $k_2 \gg k_3 \Rightarrow x \ll y$. The leading $k_2$ dependence in this limit is thus given by the integral
\be
\int d^3x \, \frac{e^{-i\vecs[k,2]\cdot\vec[x]}}{x^3} \sim ln(\lambda k_2), \, \lambda \rightarrow 0.
\label{leadk2}
\ee
Using dimensional analysis to fix the $k_3$ dependence in eq.\eqref{tpoplim}, we find that the three point function in this limit is of the form
\be
\langle O(\vecs[k,1]) O(\vecs[k,2]) O(\vecs[k,3])\rangle \sim (2\pi)^3 \, \delta^3\big(\sum_{a=1}^3 \vecs[k,a]\big) \, k_3^{\, 3} \, ln(\lambda k_2).
\label{tplim1}
\ee
From eq.\eqref{defSh}, eq.\eqref{ssimp} and eq.\eqref{tplim1},  we see that only two terms,   $(n_1,n_2) = (1,1)$ or $(-1,-1)$
are consistent with this behaviour. Now, by taking the similar limit $k_1 \ll k_2 \approx k_3$ and following the steps outlined above, we can see that the signs of $k_2$ and $k_3$ should also be identical: $(n_2,n_3) = (1,1)$ or $(-1,-1)$. Combining these two results, we see that out of the eight possibilities in eq.\eqref{ssimp} for $(n_1,n_2,n_3)$, only two survive: $(n_1,n_2,n_3) = (1,1,1)$ and $(n_1,n_2,n_3) = (-1,-1,-1)$.

Note that the choice $(n_1,n_2,n_3) = (-1,-1,-1)$ differs from $(n_1,n_2,n_3) = (1,1,1)$ only by an overall sign, which can be absorbed into the coefficient. By suitably redefining $\lambda$ and the normalization $N$ to absorb some constants, we then get $S_h$ to be given by eq.(\ref{shfinal}).


\section{A Prescription to Calculate $\langle OOO \rangle$ from $\langle OOOO \rangle$}
\label{prescription}

In this appendix, we will argue that for a given scalar four point coefficient function  $\langle OOOO \rangle$ in general, not necessarily for the canonical slow roll model, the Ward identity in eq.\eqref{wardidsct} can be solved, in principle, to get the three point coefficient function $\langle OOO \rangle$. We start by decomposing $\langle OOO \rangle'$ into two parts
\be
\langle O(\vecs[k,1]) O(\vecs[k,2]) O(\vecs[k,3]) \rangle' = S_h(k_1, k_2, k_3) + S_i(k_1, k_2, k_3)\,,
\label{sdecomp}
\ee
where $S_h(k_1, k_2, k_3)$ is the homogeneous piece eq.\eqref{shfinal}, and $S_i(k_1, k_2, k_3)$ is a particular solution to the inhomogeneous Ward identity eq.\eqref{wardidsct}. To calculate the particular solution $S_i(k_1,k_2,k_3)$, we rewrite the eq.\eqref{wardidsct} as,
\begin{align}
\begin{split}
\label{ward2g}
{\cal L}^{\bf{b}}_{\vecs[k,1]} \langle O(\vecs[k,1]) O(\vecs[k,2]) O(\vecs[k,3]) \rangle' +
{\cal L}^{\bf{b}}_{\vecs[k,2]} \langle O(\vecs[k,1]) O(\vecs[k,2]) O(\vecs[k,3]) \rangle'+
{\cal L}^{\bf{b}}_{\vecs[k,3]} \langle &O(\vecs[k,1]) O(\vecs[k,2]) O(\vecs[k,3]) \rangle' \\ &= b^j \, f_j(\vk,\vkk,\vkkk).
\end{split}
\end{align}
Here, $ f_j(\vk,\vkk,\vkkk)$ is assumed to be an arbitrary vector function of the three momenta $\vecs[k,a]$. Comparing with eq.\eqref{wardidsct}, we can see that
\be
\label{deffj}
f_j(\vk,\vkk,\vkkk) = 2 \, \frac{\dot{\bar{\phi}}}{H} \, \frac{\del}{\del k_4^j}\, \langle O(\vecs[k,1]) O(\vecs[k,2]) O(\vecs[k,3]) O(\vecs[k,4]) \rangle' \bigg\rvert_{\vecs[k,4] \rightarrow 0}.
\ee
Note that from eq.\eqref{ward2g}, $f_j(\vk,\vkk,\vkkk)$ is symmetric under the permutations of its arguments. We can write the most general vector function $f_j(\vk,\vkk,\vkkk)$ with the above property as
\be
\label{fexp}
f_j(\vk,\vkk,\vkkk) =k_{1j} \, F(k_1,k_2,k_3) + k_{2j} \, F(k_2,k_3,k_1) + k_{3j} \, F(k_3,k_1,k_2)\,,
\ee
such that $F(k_1,k_2,k_3)$ is an arbitrary function and is symmetric under the exchange of its last two arguments.

Next, we make a choice for $\vec[b]$, the parameter of special conformal transformation, to be perpendicular to $\vkkk$,
\be
\label{choiceb1}
\vec[b] = \vkk -{\vkk \cdot \vkkk \over k_3^2} \vkkk .
\ee
Using eq.\eqref{fexp} and eq.\eqref{choiceb1}, the RHS of eq.\eqref{ward2g} becomes
\be
\label{ward2grhs}
b^j \, f_j(\vk,\vkk,\vkkk) = \bigg(k_2^2 - {(\vkk \cdot \vkkk)^2 \over k_3^2}\bigg) \, g(k_1,k_2,k_3),
\ee
with the definition,
\be
\label{defg}
g(k_1,k_2,k_3)=F(k_2,k_1,k_3)-F(k_1,k_2,k_3).
\ee
It is obvious from the definition that $g(k_1,k_2,k_3)$ is antisymmetric under the exchange of its first two arguments.
Also, using eq.\eqref{deltao4}, eq.\eqref{deftheta} and eq.\eqref{choiceb1}, we can write the LHS of eq.\eqref{ward2g} as,
\begin{align}
\begin{split}
{\cal L}^{\bf{b}}_{\vecs[k,1]} \langle O(\vecs[k,1]) O(\vecs[k,2]) O(\vecs[k,3]) \rangle' +
{\cal L}^{\bf{b}}_{\vecs[k,2]} \langle O(\vecs[k,1]) O(\vecs[k,2]) O(\vecs[k,3]) \rangle'+
{\cal L}^{\bf{b}}_{\vecs[k,3]} \langle O(\vecs[k,1]) O(\vecs[k,2]) O(\vecs[k,3]) \rangle' \\ = \bigg(k_2^2 - {(\vkk \cdot \vkkk)^2 \over k_3^2}\bigg) (\Theta(k_2)-\Theta(k_1)) \, S_i(k_1, k_2, k_3).
\label{ward2glhs1}
\end{split}
\end{align}
From eq.\eqref{ward2grhs} and eq.\eqref{ward2glhs1}, we see that the Ward identity eq.\eqref{ward2g} becomes,
\be
\label{wardg1}
(\Theta(k_2)-\Theta(k_1)) \, S_i(k_1, k_2, k_3) = g(k_1,k_2,k_3).
\ee
Next we expand both $S_i(k_1, k_2, k_3)$ and $g(k_1,k_2,k_3)$ in terms of the functions $f_z(k)$, eq.\eqref{deffzk},
\begin{align}
\label{exps}
S_i(k_1,k_2,k_3)=  \int_{-\infty}^{\infty} dz_1 \,dz_2 \,dz_3 \, \mathcal{F}(k_1,k_2,k_3,z_1,z_2,z_3) \, \mathcal{M}(z_1,z_2,z_3), \\
\label{expg}
g(k_1,k_2,k_3) = \int_{-\infty}^{\infty} dz_1 \,dz_2 \,dz_3 \, \mathcal{F}(k_1,k_2,k_3,z_1,z_2,z_3) \, \mathcal{N}(z_1,z_2,z_3),
\end{align}
with
\be
\mathcal{F}(k_1,k_2,k_3,z_1,z_2,z_3) = (1 + i k_1 z_1)\,(1 + i k_2 z_2)\,(1 + i k_3 z_3)\,e^{-i (k_1 z_1+ k_2z_2+ k_3z_3)}.
\label{funcf1}
\ee
Substituting eq.\eqref{exps} and eq.\eqref{expg} into eq.\eqref{wardg1} gives us,
\be
\label{relmn}
\mathcal{M}(z_1,z_2,z_3) = {\mathcal{N}(z_1,z_2,z_3) \over z_1^2 - z_2^2} .
\ee
Using the definition of the inverse transformation in eq.\eqref{invsf}, we can invert eq.\eqref{expg} to obtain $\mathcal{N}(z_1,z_2,z_3)$ in terms of $g(k_1,k_2,k_3)$ as
\begin{align}
\begin{split}
\label{invgn}
\mathcal{N}(z_1,z_2,z_3) = -\int_{-\infty}^{\infty} {dk_1 \over 2\pi}{dk_2 \over 2\pi}{dk_3 \over 2\pi} & \, k_1 \, k_2 \, k_3 \, e^{i (k_1 z_1+ k_2z_2+ k_3z_3)} \\ & \bigg(\int^{k_1} \int^{k_2}\int^{k_3} {g(q_1,q_2,q_3) \over q_1^2 q_2^2 q_3^2} \, dq_1 dq_2 dq_3 \bigg).
\end{split}
\end{align}
Using eq.\eqref{invgn} and eq.\eqref{relmn} in eq.\eqref{exps}, we finally obtain
\begin{align}
\begin{split}
\label{solinhomo}
S_i(k_1,k_2,k_3) = -\int_{-\infty}^{\infty} d&z_1 dz_2 dz_3 \, {\mathcal{F}(k_1,k_2,k_3,z_1,z_2,z_3) \over (z_1^2 - z_2^2)} \, \bigg[ \int_{-\infty}^{\infty} {dp_1 \over 2\pi}{dp_2 \over 2\pi}{dp_3 \over 2\pi} \, p_1 p_2 p_3 \\ & e^{i (p_1 z_1+ p_2 z_2+ p_3 z_3)}
\bigg(\int^{p_1} \int^{p_2} \int^{p_3} {g(q_1,q_2,q_3) \over q_1^2q_2^2q_3^2} \, dq_1dq_2dq_3 \bigg) \bigg].
\end{split}
\end{align} 

Thus, given a four point coefficient function $\langle OOOO \rangle$, we should first calculate the function $g(q_1,q_2,q_3)$, eq.\eqref{defg}. Knowing $g$, we can evaluate the integral in eq.\eqref{solinhomo} to obtain the function $S_i$. Eq. \eqref{sdecomp} then gives us the three point coefficient function $\langle OOO \rangle$, as desired.
Note that the expression above is a formal one. In particular, we know that the solution to the Ward identities is not unique, with an ambiguity of 
 the form given by $S_h$, eq.(\ref{shfinal}). This ambiguity should be related to an ambiguity in how to carry out the integrals in eq.(\ref{solinhomo}).

\bibliography{references}{}
\bibliographystyle{JHEP}
\end{document}